\documentclass[12pt,a4paper]{article}

\usepackage{graphicx}
\usepackage{amsfonts}
\usepackage{amsmath}
\usepackage{hyperref}
\usepackage{psfrag,subfigure,epsfig}
\usepackage{color,colordvi}
\usepackage{tikz}

\usepackage[numbers,square,comma,compress]{natbib} 

\csname @addtoreset\endcsname{equation}{section}

\usepackage[lmargin=1.5 cm, rmargin=1.5 cm, tmargin=3.0cm, bmargin=3.0cm]{geometry}

\usepackage{float}

\newcommand{\bi}{\bar{\imath}}
\newcommand{\bI}{\bar{I}}
\newcommand{\bj}{\bar{\jmath}}
\newcommand{\bJ}{\bar{J}}
\newcommand{\ins}{\bar{\phi}_{\star}}
\newcommand{\hins}{\hat{\phi}_{\star}}
\newcommand{\inde}{\star}

\newcommand{\mc}{\mathcal}
\newcommand{\mb}{\mathbb}

\title{\begin{flushright}{\vspace{-2.4cm}\small CERN-PH-TH/2015-159	}\end{flushright}
\vspace{0.8cm}
\bf{Probing the Moduli Dependence of Refined Topological Amplitudes}\\[20pt]}
\author{\Large  I.~Antoniadis\footnote{{\tt antoniad@lpthe.jussieu.fr}}~,~I. Florakis\footnote{\tt florakis.ioannis@cern.ch}~,~ 
S. Hohenegger\,\footnote{\tt s.hohenegger@ipnl.in2p3.fr}~,\\[0.2cm] \Large K.S.~Narain\footnote{{\tt narain@ictp.trieste.it}}~ and A.~Zein Assi\footnote{\tt zeinassi@cern.ch}}
\date{}

\begin{document}

\maketitle

\begin{center}
\renewcommand{\thefootnote}{\fnsymbol{footnote}}\vspace{-0.5cm}
${}^{\footnotemark[1]}$ Albert Einstein Center for Fundamental Physics, Institute for Theoretical Physics, University of Bern, 5 Sidlestrasse, CH-3012 Bern, Switzerland\\[0.2cm]
${}^{\footnotemark[1]}$LPTHE, UMR CNRS 7589, Sorbonne Universit\'es, UPMC Paris 6, 75005 Paris, France\\[0.2cm]
${}^{\footnotemark[2]}$ Department of Physics, CERN - Theory Division, CH-1211 Geneva 23, Switzerland\\[0.2cm]
${}^{\footnotemark[3]}$ Universit\'e Claude Bernard (Lyon 1)\\UMR 5822, CNRS/IN2P3, Institut de Physique Nucl\'eaire, Bat. P. Dirac\\ 4 rue Enrico Fermi, F-69622-Villeurbanne, France\\[0.2cm]
${}^{\footnotemark[4]\footnotemark[5]}$ High Energy Section, The Abdus Salam International Center for Theoretical Physics,
\\Strada Costiera, 11-34014 Trieste, Italy\\[1.3cm]
\end{center}

\abstract{With the aim of providing a worldsheet description of the refined topological string, we continue the study of a particular class  of higher derivative couplings $F_{g,n}$ in the type II string effective action compactified on a Calabi-Yau threefold. We analyse first order differential equations in the anti-holomorphic moduli of the theory, which relate the $F_{g,n}$ to other component couplings. From the point of view of the topological theory, these equations describe the contribution of non-physical states to twisted correlation functions and encode an obstruction for interpreting the $F_{g,n}$ as the free energy of the refined topological string theory. We investigate possibilities of lifting this obstruction by formulating conditions on the moduli dependence under which the differential equations simplify and take the form of generalised holomorphic anomaly equations. We further test this approach against explicit calculations in the dual heterotic theory.}

\newpage

\tableofcontents

%%%%%%%%%%%%%%%%%%%%%%

\section{Introduction}
Since the first construction of topological string theory \cite{Witten}, its  connection to higher derivative couplings in the string effective action has been a very active and fruitful field of study. Indeed, in \cite{Antoniadis:1993ze}, a series of higher loop scattering amplitudes $F_g$, in type II string theory compactified on a Calabi-Yau threefold, was computed and shown to capture the genus $g$ free energy of the topological string. These couplings are BPS protected and involve $2g$ chiral supergravity multiplets. The result of \cite{Antoniadis:1993ze} is interesting from a number of different perspectives. On the one hand, the $F_g$ encode very important target space physics, for example in computing macroscopic corrections to the entropy of supersymmetric black holes (see for example~\cite{BlackHole}). On the other hand, they provide a concrete worldsheet description of the topological string which is very powerful in studying its properties \cite{Bershadsky:1993cx}.

During the last two decades, the work of \cite{Antoniadis:1993ze} has been extended and many new relations between topological correlation functions and higher derivative effective couplings in string theory have been found~\cite{Berkovits:1994vy,Berkovits:1998ex,Ooguri:1995cp,Antoniadis:1996qg,Antoniadis:2005sd,Antoniadis:2006mr,Antoniadis:2007cw,Antoniadis:2009nv,Antoniadis:2011hq}. Along these lines, it was suggested in \cite{Antoniadis:2010iq} that a suitable generalisation $F_{g,n}$ of the $F_g$ could provide a worldsheet description of the refined topological string. The refinement of the topological string consists of a one-parameter deformation of topological string theory, inspired by recent progress in the study of supersymmetric gauge theories~\cite{Moore:1997dj,Lossev:1997bz,Nekrasov:2002qd}, so that its point-particle limit reproduces the partition function of supersymmetric gauge theories in the full $\Omega$-background. In this correspondence, the topological string coupling $g_s$ is 
identified with one of the geometric deformation parameters $\epsilon_-$, while the refinement is an extension associated to the second parameter $\epsilon_+$. The first proposal  successfully satisfying this requirement was presented in \cite{Antoniadis:2013bja}, through explicit computations to all orders in $\alpha'$ in heterotic string theory.

From the target space point of view, numerous different descriptions of the refinement exist, such as the counting of particular BPS-states in M-theory \cite{Mtheory}, the refined topological vertex \cite{Topvertex}, matrix models using refined ensembles \cite{MatrixTheory} or through a construction of the $\Omega$-background using the so-called flux-trap background \cite{Hellerman:2011mv}. In a recent work \cite{Antoniadis:2013bja}, we proposed a worldsheet description of the refined topologic string using a generalisation of the couplings $F_{g}$ involving two Riemann tensors and $2g-2$ insertions of graviphoton field strengths, by additional insertions of chiral projections of specific vector multiplets. These couplings are of the general form discussed in \cite{Morales:1996bp,Antoniadis:2010iq} (see also \cite{Antoniadis:1993ze,Antoniadis:1996qg,Antoniadis:2007cw,Antoniadis:2009nv,Antoniadis:2011hq}).  The precise nature of the additional insertions is crucial in exactly reproducing the Nekrasov partition 
function both perturbatively \cite{Antoniadis:2013bja} and non-perturbatively \cite{Antoniadis:2013mna}. Specifically, working in heterotic string theory compactified on $K3\times T^2$, we computed in \cite{Antoniadis:2013bja} a series of refined couplings $F_{g,n}^{\bar{T}}$ which include additional $2n$ insertions of the field strength tensor of the vector superpartner of the K\"ahler modulus of $T^2$ ($\bar{T}$-vector). These amplitudes are exact to all orders in $\alpha' $ and start receiving corrections at the one-loop level in $g_s$. At a particular point of enhanced gauge symmetry in the heterotic moduli space, they reproduce exactly the perturbative part of the Nekrasov partition function in the point particle limit for arbitrary values of the deformation parameters.\footnote{See \cite{Nakayama:2011be} for a different proposal reproducing the Nekrasov partition function to leading order.} A very strong check of our proposal was performed in~\cite{Antoniadis:2013mna} (see \cite{Assi:2014exa,Florakis:2014gja} for reviews and \cite{Billo:2006jm,Ito:2010vx} for earlier partial results) by computing gauge theory instanton corrections to $F_{g,n}$, which precisely reproduce also the non-perturbative part of the gauge theory partition function.

The connection between the couplings studied in \cite{Antoniadis:2013bja} and the full Nekrasov partition function is a very strong hint that our proposal for the $F_{g,n}^{\bar T}$ can indeed furnish a worldsheet description of the refined topological string. In this context, non-physical states of the topological theory are required to decouple from $F_{g,n}$. In the unrefined case (\emph{i.e.} for $n=0$), this requirement has first been studied in~\cite{Bershadsky:1993cx}: in the twisted theory, the BRST operator is identified with one of the supercharges of the original $\mathcal{N}=2$ worldsheet superconformal theory. Thus, some of the moduli of the untwisted theory are not part of the topological BRST cohomology and are `unphysical' from the latter point of view. This implies that $F_{g}$ should possess holomorphy properties. In the supergravity formulation, this agrees with the fact that the $F_g$ only depend on the chiral vector multiplet moduli and can be written in the form of BPS-saturated F-
terms in $\mathbb{R}^{4|8}$ superspace. However, as pointed out in \cite{Bershadsky:1993cx}, in string theory, there is a residual dependence on the anti-holomorphic moduli due to boundary effects in the moduli space of the higher genus worldsheet. This gives rise to a recursive differential equation known as the \emph{holomorphic anomaly equation}, which relates the anti-holomorphic moduli derivative of $F_{g}$ to combinations of (holomorphic derivatives of) $F_{g'}$ with $g'<g$. 

In this paper we study the question of the decoupling of anti-holomorphic moduli in the case of the $F_{g,n}$ studied in \cite{Antoniadis:2010iq,Antoniadis:2013bja} by deriving differential equations for the corresponding effective couplings. For $n>0$, the $F_{g,n}$ are no longer F-terms, but also contain chiral projections of superfields. Therefore, \emph{a priori}, there are no constraints on their dependence on anti-holomorphic moduli, even at the level of supergravity. However, by analysing the structure of the couplings in superspace, we obtain differential equations which relate anti-holomorphic derivatives of $F_{g,n}$ to new component couplings, and the latter can be realised as scattering amplitudes in string theory. 

By studying these relations in detail in supergravity, we can reformulate the vanishing of the anti-holomorphic vector multiplet dependence in $F_{g,n}$ as well-defined conditions on the moduli dependence of particular coupling functions in the effective action. The latter conditions go beyond the constraints of $\mathcal{N}=2$ supersymmetry and might be interpreted as a consequence of a $U(1)$ isometry present in a special region in the string moduli space, as required from the point of view of gauge theory in order to formulate a supersymmetric $\Omega$-background \cite{Moore:1997dj,Lossev:1997bz,Nekrasov:2002qd}. In this case, since such isometries are generically not present in compact Calabi-Yau threefolds, the conditions for decoupling the anti-holomorphic vector multiplets might be regarded as Ward identities related to the appearance of $U(1)$ isometries in suitable decompactification limits.

Extending the supergravity analysis, we derive explicit differential equations for the $F_{g,n}$ in the framework of the fully-fledged type II string theory compactified on generic Calabi-Yau threefolds. We relate all new component couplings involved in these relations in the form of higher genus scattering amplitudes and express them as twisted worldsheet correlators on a genus $g$ Riemann surface with $2n$ punctures. The equations we obtain contain corrections induced by boundary effects in the moduli space of the higher genus worldsheet. From the string theory perspective, the decoupling of non-holomorphic moduli translates into well-defined conditions on the worldsheet correlators. The upshot of our approach is that it provides a solid framework, based on physical string couplings, in which the above mentioned Ward identities may be analysed in the full worldsheet theory. In particular, we can formulate conditions under which the string-derived differential equations reduce to the recursive structure of a generalised holomorphic anomaly equation. Equations of this type were postulated in \cite{Huang:2010kf,Huang:2011qx} as the definition of the refined topological string.

Finally, we also study the differential equations in the dual setup of heterotic string theory on $K3\times T^2$. On the heterotic side, the $F_{g,n}^{\bar{T}}$ start receiving contributions at the one-loop level and therefore constitute the ideal testing-ground for the ideas developed in type II, particularly for certain decompactification limits. We find that in the large volume limit of $T^2$, they satisfy recursive differential equations which precisely match with the weak coupling version of our differential equations in type II, hence providing a non-trivial check of our approach. On the other hand, we use the heterotic setup to study boundary conditions to the differential equations developed in this work. Indeed, in \cite{Huang:2010kf,Huang:2011qx}, the field theory limit was used as a boundary condition to solve for the couplings $F_{g,n}$. In the present case, while the equations in type II are essentially covariant with respect to the choice of vector multiplet insertion in $F_{g,n}$, only the specific choice of the $\bar{T}$-vector for $F_{g,n}^{\bar{T}}$ was found in \cite{Antoniadis:2013bja} to reproduce the gauge theory partition function. Here, we show that also other choices of vector multiplet insertions lead to the same boundary conditions when expanded around an appropriate point of enhanced gauge symmetry in the heterotic moduli space.

The paper is organised as follows. In Section~\ref{Sec:Sugra}, we prepare the ground by discussing the effective action couplings $F_{g,n}$ and extract several relations among them implied by supersymmetry. In Section~\ref{Sec:TypeII}, we derive equations in type II string theory compactified on a Calabi-Yau threefold. We derive all necessary amplitudes at higher genus and identify string theoretic corrections to the supergravity equations as boundary terms of the worldsheet moduli space. In Section~\ref{Sec:LocalLimit}, we discuss simplifications of the differential equations which we propose to be the effect of $U(1)$ isometries of the target space Calabi-Yau threefold. In particular, we point out that, under certain conditions, a recursive structure emerges in the equations, both at the supergravity and at the full string level in type II. In Section~\ref{Sect:HetRealisation}, we consider the dual heterotic theory on $K3\times T^2$. We first perform a check of the results obtained in type II from the heterotic dual computation and then provide boundary conditions to the differential equations by reproducing the Nekrasov partition function for different vector multiplet insertions in $F_{g,n}$. Finally, Section~\ref{Sec:Conclusions} contains a summary of our results and our conclusions. Several technical results are compiled in three appendices.
%%%%%%%%%%%%%%%%%%%%%%%%%%%%%%%%%%%%%%%%%%%%%%

\section{String Effective Couplings}\label{Sec:Sugra}
The central object of this paper is a particular class of higher-derivative effective couplings in $\mathcal{N}=2$ supersymmetric string compactifications to four dimensions, which were considered in \cite{Morales:1996bp,Antoniadis:2010iq} (see also \cite{Antoniadis:1993ze,Antoniadis:1996qg,Antoniadis:2007cw,Antoniadis:2009nv,Antoniadis:2011hq}) in the form of \emph{generalised F-terms}. In this section, we demonstrate that supersymmetry requires a number of consistency relations among different component couplings.
%%%%%%%%%%%%%%%%%%%%%%%%%%%%%%%%%%%%%%%%%%%%%%
\subsection{Superspace Description}\label{Sect:Couplings}
We begin by reviewing the general class of string effective component couplings   \cite{Antoniadis:2010iq}  of the form
\begin{align}
\int d^4x\,  F_{g,n}^{\,I_1\ldots I_{2n}}(\varphi,\bar{\varphi})\,R_{(-)}^2 \left(F_{(-)}^G\right)^{2g-2} \left(F_{(+)}^{I_1} \cdot F_{(+)}^{I_2}\right)\ldots \left(F_{(+)}^{I_{2n-1}} \cdot F_{(+)}^{I_{2n}}\right)\,, \label{RefinedCouplings}
\end{align}
where $R_{(-)}$ is the (self-dual) Riemann tensor, $F^G_{(-)}$ the (self-dual) graviphoton field-strength tensor and $F_{(+)}^{I}$ the (anti-self-dual) field strength tensor of a physical vector multiplet gauge field $A^I_\mu$, which we label by the index $I$, with $I=1,\ldots,N_V$, and $\mu$ is a space-time Lorentz index. In general, the coupling function $F_{g,n}^{\,I_1\ldots I_{2n}}$ depends covariantly on the vector multiplet moduli in a non-holomorphic fashion. Only the case $n=0$ is special, for which (\ref{RefinedCouplings}) reduces to a series of holomorphic couplings \cite{Antoniadis:1993ze} of the vector multiplet moduli.

The supersymmetric version of the component terms (\ref{RefinedCouplings}) can be described in standard superspace $\mathbb{R}^{4|8}$ parametrised by the coordinates $(x_\mu,\theta_\alpha^a,\bar{\theta}_a^{\dot{\alpha}})$. 
To this end, we introduce the $\mathcal{N}=2$ supergravity multiplet
\begin{align}
W_{\mu\nu}^{ab}=\epsilon^{ab}\left(F^{G}_{(-)}\right)_{\mu\nu}+\ldots+(\theta^a\sigma^{\rho\tau}\theta^b)R_{(-)\mu\nu\rho\tau}
\end{align}
as well as the chiral and anti-chiral vector multiplets
\begin{align}
&X^I=\varphi^I+\theta_a^\alpha\,\lambda_\alpha^{Ia}+\epsilon_{ab}(\theta^a\sigma^{\mu\nu}\theta^b)\,F^I_{(-)\mu\nu}+\ldots\,,\\
&\bar{X}^I=\bar{\varphi}^I+\bar{\theta}_{\dot{\alpha}}^a\,\bar{\lambda}_{a}^{I\dot{\alpha}}+\epsilon^{ab}(\bar{\theta}_a\bar{\sigma}^{\mu\nu}\bar{\theta}_b)\,F^I_{(+)\mu\nu}+\ldots\,.
\end{align}
In addition, we define the descendent fields
\begin{align}
\bar{K}_{\mu\nu}^I=\left(\epsilon_{ab}\bar{D}^a\bar{\sigma}_{\mu\nu}\bar{D}^b\right) \bar{X}^I=F_{(+)\mu\nu}^I+\ldots\,,
\end{align}
where $\bar{D}^a_{\dot{\alpha}}$ are the (anti-)chiral spinor derivatives. On-shell, these descendents are chiral objects in the sense that
\begin{align}
\bar{D}^i_{\dot{\alpha}}\bar{K}^I_{\mu\nu}=0\,.\label{ChiralDescendent}
\end{align}
We can use these superfields to write a superspace version of the component couplings (\ref{RefinedCouplings}):
\begin{align}
\int d^4x\, d^4\theta\,\left(\bar{D}^a\bar{\sigma}_{\mu\nu}\bar{D}_a\right)^2\left[\mathbb{F}^{I_1\ldots I_{2n-2}}_{g,n}(X,\bar{X})\,(W_{\mu\nu}^{ab}W^{\mu\nu}_{ab})^g\,(\bar{K}^{I_1}\cdot \bar{K}^{I_2})\ldots (\bar{K}^{I_{2n-3}}\cdot \bar{K}^{I_{2n-2}})\right]\,.\label{GeneralDterm}
\end{align}
The non-holomorphic coefficient functions $\mathbb{F}^{I_1\ldots I_{2n-2}}_{g,n}(X^I,\bar{X}^I)$ in (\ref{GeneralDterm}) are generic symmetric tensors transforming in some (reducible) representation of the T-duality group. Upon expansion in the Grassmann variables they can be related to coefficient couplings, which in turn are related to scattering amplitudes that we study in Section~\ref{Sec:TypeII} in type II string theory.
%%%%%%%%%%%%%%%%%%%%%%%%%%%%%%%%%
\subsection{Component Couplings and Differential Equations}\label{Sect:SugraDiffEqu}
In (\ref{GeneralDterm}), all vector multiplets have been treated on an equal footing and we have considered the couplings $F_{g,n}^{I_1\ldots I_{2n}}$ as generic tensors of the $SO(N_V)$ T-duality group. In the following, we focus on couplings involving only one singled out vector multiplet, which we denote by $(X^\star,\bar{X}^\star)$. According to our proposal \cite{Antoniadis:2013bja} for a worldsheet description of the  refined topological string, $\bar{\varphi}^\star$ should be identified with the K\"ahler modulus of the torus in the heterotic $K3\times T^2$ compactification. Here, however, we do not yet give a particular geometric interpretation of $X^\star$ pertaining to a specific model and we keep the discussion general.

Concerning our notation, we introduce the indices $i,j$ (as well as $\bi,\bj$) which run over all vector multiplets except $(X^\star,\bar{X}^\star)$. We also utilise the notation
\begin{align}
&\mathbb{F}_{g,n}:=\mathbb{F}^{\star\ldots\star}_{g,n}\,,&&\mathbb{F}^{\bi}_{g,n}:=\mathbb{F}^{\bi,\star\ldots\star}_{g,n}\,,&&\mathbb{F}^{\bi\bj}_{g,n}:=\mathbb{F}^{\bi\bj,\star\ldots\star}_{g,n}\,,&&\text{etc.}
\end{align}
In order to extract component couplings from the superspace expression (\ref{GeneralDterm}), we perform the anti-chiral spinor derivatives as well as the integration over the chiral Grassmann variables. We focus on component terms that contain two (self-dual) Riemann tensors and $2g-2$ (self-dual) graviphoton field strength tensors. Therefore, we consider 
\begin{align}
(W_{\mu\nu}^{ab}W^{\mu\nu}_{ab})^g\sim \theta_\alpha^a\,\theta^\alpha_b\,\theta^b_\beta\,\theta^\beta_a\,R_{(-),\,\mu\nu\rho\tau}\,R^{\,\mu\nu\rho\tau}_{(-)}\,\left[F^G_{(-)}\cdot F^{(G)}_{(-)}\right]^{g-1}+\ldots\,,
\end{align}
and use the leading term to saturate the chiral theta integration, such that it remains to distribute the anti-chiral spinor derivatives and choose the contribution $\bar{\theta}_a^{\dot{\alpha}}=0$ in the end. We recall that the anti-chiral vector multiplets contain only the anti-chiral spinor components $\bar{\lambda}^{A,a}_{\dot{\alpha}}$ as well as the anti-self dual part of the gauge field strength tensor $F^A_{(+)\mu\nu}$. For the latter, we denote the three independent components as $\{F^{(++)},F^{(0)},F^{(--)}\}$, which are labelled by the charges with respect to the anti-self-dual $SU(2)\subset SO(4)$ subgroup of the Lorentz group. In particular, we have
\begin{align}
(\bar{K}_{I_1}\cdot \bar{K}_{I_2})=F_{I_1}^{(++)}F_{I_2}^{(--)}+F_{I_1}^{(0)}F_{I_2}^{(0)}+F_{I_1}^{(--)}F_{I_2}^{(++)}+\ldots\,,
\end{align}
where the dots denote higher terms in the Grassmann variables. Furthermore, we also find at the component level
{\allowdisplaybreaks
\begin{align}
&(\bar{\mathcal{D}}_\star^2\,\mathbb{F}_{g,n})(\bar{K}_\star\cdot \bar{K}_\star)^{n-1}=T^{(1)}_{g,n}\,(F_\star^{(++)})^n\,(F_\star^{(--)})^n+\ldots\,,\nonumber\\
&(\bar{\mathcal{D}}_\star^3\,\mathbb{F}_{g,n})(\bar{K}_\star\cdot \bar{K}_\star)^{n-1}=T^{(2)}_{g,n}\,(F_\star^{(++)})^n\,(F_\star^{(--)})^{n-1}F_{\star}^{(--)}(\bar{\lambda}^-_\star\cdot \bar{\lambda}^-_\star)+\ldots\,,\nonumber\\
&(\bar{\mathcal{D}}_\star \bar{\mathcal{D}}_{\bi}\,\mathbb{F}_{g,n})(\bar{K}_\star\cdot \bar{K}_\star)^{n-1}+(\bar{\mathcal{D}}_\star^2\,\mathbb{F}_{g,n;\bi})(\bar{K}_\star\cdot \bar{K}_\star)^{n-2}(\bar{K}_\star\cdot \bar{K}_{\bi})\nonumber\\
&\hspace{3cm}=T_{g,n;\bi}^{(1)}\,(F_\star^{(++)})^n\,(F_\star^{(--)})^{n-1}\,F_{\bi}^{(--)}+\ldots\,,\nonumber\\
&(\bar{\mathcal{D}}_\star^3\,\mathbb{F}_{g,n;\bi})(\bar{K}_\star\cdot \bar{K}_\star)^{n-2}(\bar{K}_\star\cdot \bar{K}_{\bi})=T_{g,n;\bi}^{(2)}\,(F_\star^{(++)})^n\,(F_\star^{(--)})^{n-2}\,F_{\bi}^{(--)}(\bar{\lambda}^-_\star\cdot \bar{\lambda}^-_\star)+\ldots\,,
\end{align}}
where the dots denote additional terms (including $\bar{\theta}_a^{\dot{\alpha}}$) that we are not be interested in, and $\mc D_I,\bar{\mc D}_{\bar I}$ are holomorphic, anti-holomorphic K\"ahler covariant derivatives. In addition, we have introduced
\begin{align}
&T^{(1)}_{g,n}=\bar{\mathcal{D}}_\star^2\,\mathbb{F}_{g,n}\big|_{\theta=\bar{\theta}=0}\,,&&T^{(2)}_{g,n}=\bar{\mathcal{D}}_\star^3\,\mathbb{F}_{g,n}\big|_{\theta=\bar{\theta}=0}\,,\nonumber\\
&T_{g,n;\bi}^{(1)}=(\bar{\mathcal{D}}_\star \bar{\mathcal{D}}_{\bi}\,\mathbb{F}_{g,n}+\bar{\mathcal{D}}_\star^2\,\mathbb{F}_{g,n;\bi})\big|_{\theta=\bar{\theta}=0}\,,&&T_{g,n;\bi}^{(2)}=\bar{\mathcal{D}}_\star^3\,\mathbb{F}_{g,n;\bi}\big|_{\theta=\bar{\theta}=0}\,.\label{DefTs}
\end{align}
We can relate these quantities to explicit scattering amplitudes in the effective action:
\begin{align}
F_{g,n}&=(n!)^2\,T^{(1)}_{g,n}\,=\left\langle (R_{(-)}\cdot R_{(-)})(F^G_{(-)}\cdot F^G_{(-)})^{g-1}\,(F_\star^{(++)})^n\,(F^{(--)}_\star)^n\right\rangle\,,\nonumber\\
\Psi_{(\star\star|\star)}^{g,n}&=n!(n-1)!\,T_{g,n}^{(2)}=\left\langle (R_{(-)}\cdot R_{(-)})(F^G_{(-)}\cdot F^G_{(-)})^{g-1}\,(F_\star^{(++)})^n(F_\star^{(--)})^{n-1}\,(\bar{\lambda}^-_\star\cdot \bar{\lambda}^-_\star)\right\rangle\,,\nonumber\\
F_{g,n,\bi}&=n!(n-1)!\,T_{g,n;\bi}^{(1)}\,=\left\langle (R_{(-)}\cdot R_{(-)})(F^G_{(-)}\cdot F^G_{(-)})^{g-1}\,(F_\star^{(++)})^n\,(F_\star^{(--)})^{n-1}\, F_{\bi}^{(--)}\right\rangle\,,\nonumber\\
\Psi_{(\star\star|\bi)}^{g,n}&=n!(n-2)!\,T_{g,n;\bi}^{(2)}\,\nonumber\\
&=\left\langle (R_{(-)}\cdot R_{(-)})(F^G_{(-)}\cdot F^G_{(-)})^{g-1}\,(F_\star^{(++)})^n(F_\star^{(--)})^{n-2}\, F_{\bi}^{(--)}\,(\bar{\lambda}^-_\star\cdot \bar{\lambda}^-_\star)\right\rangle\,.\label{AmpsDefEffAction}
\end{align}
From the definitions (\ref{DefTs}), we deduce that\footnote{There are several other identities that we can find in this manner, however, in the remainder of this work, we only study (\ref{RelSUGRA0}) and (\ref{RelSUGRA}).}
\begin{align}
\bar{\mathcal{D}}_{\star}T^{(1)}_{g,n}&=T_{g,n}^{(2)}\,,\label{RelSUGRA0}\\
\bar{\mathcal{D}}_{\bi}T^{(1)}_{g,n}&=\bar{\mathcal{D}}_\star\,T_{g,n;\bi}^{(1)}-T_{g,n;\bi}^{(2)}\,,\label{RelSUGRA}
\end{align}
which translate into the following relations for the amplitudes:
\begin{align}
\bar{\mathcal{D}}_\star F_{g,n}&=n\,\Psi^{g,n}_{(\star\star|\star)}\,,\label{bWEquPredict0}\\
\bar{\mathcal{D}}_{\bi}\,F_{g,n}&=n\,\bar{\mathcal{D}}_\star\,F_{g,n,\bi}-n(n-1)\Psi^{g,n}_{(\star\star|\bi)}\,.\label{bWEquPredict}
\end{align}
The latter are a consequence of supersymmetry and the particular structure of the effective couplings (\ref{GeneralDterm}). 
%%%%%%%%%%%%%%%%%%%%%%%%%%%%%%%%%%%%%%%%%%%%%%%%%%%

\section{Differential Equations in Type II}\label{Sec:TypeII}
%%%%%%%%%%%%%%%%%%%%%%%%%%%%%%%%%%%%%%%%%%%%%%%%%%%
\subsection{\texorpdfstring{Type II Genus $g$ Amplitudes}{Type II Genus g Amplitudes}}\label{Sec:GeneralTypeII}
In this section, we consider realisations of the couplings (\ref{AmpsDefEffAction}) discussed above as genus-$g$ string amplitudes in type II string theory on a generic Calabi-Yau threefold $X$ and derive generalisations of the relations (\ref{bWEquPredict0}) and (\ref{bWEquPredict}) in the framework of fully-fledged string theory. 

\subsubsection{Gauge Field Amplitudes}
We begin by providing an expression for the $F_{g,n,\bar{I}}$ and then proceed to consider the differential equations they satisfy. The key ingredient to deriving the couplings $F_{g,n}$ is the vertex operator of the (anti-self-dual) vector multiplet gauge field strength tensor $F_\star$. In the $-\tfrac{1}{2}$ ghost picture, it takes the form
\begin{align}
V^{(-1/2)}_{\star}(z,\bar{z})=\eta^\mu p ^\nu e^{-\frac{1}{2}(\hat{\varphi}+\tilde{\hat{\varphi}})}\,(S \bar{\sigma}_{\mu\nu} \tilde{S})\,\Sigma_{\star}\, (z,\bar{z})\,e^{ip\cdot Z}\,,\label{VertexDef}
\end{align}
where $z$ is the insertion point of the vertex on the worldsheet, $\hat{\varphi}$ ($\tilde{\hat{\varphi}}$) are the left-(right-)moving super ghost fields and $S$ ($\tilde{S}$) are the left-(right-)moving space-time spin fields. Furthermore, $\eta^\mu$ and $p^\mu$ are the polarisation and space-time momentum respectively (which satisfy $\eta\cdot p=0$) and $Z^\mu$ are the space-time coordinates. The nature of the vector multiplet is determined by the internal field $\Sigma_{\star}$. Concretely, upon bosonising the $U(1)$ Kac-Moody currents $J$ and $\tilde{J}$ in terms of $H$ and $\tilde{H}$ respectively, we can write
\begin{align}
\Sigma_{\star}(z,\bar{z})=\lim_{w\to z} |w-z|\,e^{\frac{\sqrt{3}\,i}{2}\,\left(H(w)\mp\tilde{H}(\bar{w})\right)}\,\bar{\phi}_{\star}(w,\bar{w})\,,\label{VertexInternal}
\end{align}
where $\bar{\phi}_{\star}$ is an (anti-chiral,chiral) primary ((anti-chiral,anti-chiral) primary) state of the type IIA (type IIB) worldsheet theory. Assuming that the vector multiplet gauge fields $A^\mu_\star$ have no contact terms among themselves\footnote{Specific conditions for this to happen have been formulated in \cite{Antoniadis:2010iq}.}, which would require the subtraction of 1PI reducible diagrams, the $g$-loop amplitude can be written \cite{Antoniadis:2010iq} in the form of a twisted worldsheet correlator integrated over the moduli space $\mathcal{M}_{g,n}$ of a genus $g$ Riemann surface $\Sigma_{g,n}$ with $n$ punctures (located at positions $u_\ell$):
\begin{align}
F_{g,n}=\int_{\mathcal{M}_{g,n}}\left\langle\prod_{k=1}^{3g-3+n}|\mu_k\cdot G^-|^2\left(\prod_{m=1}^n\int_{\Sigma_{g,n}} \ins (z_m)\right)\left(\prod_{\ell=1}^n\hins(u_\ell)\right)\right\rangle_{\text{twist}}\,.\label{TwistAmplitude}
\end{align}
Our notation allows us to treat type IIA and type IIB string theory simultaneously. Indeed, for the measure, we use the shorthand 
\begin{align}
|\mu\cdot G^-|^2:=\left\{\begin{array}{lcl}G^-(\mu)\,\tilde{G}^+(\bar{\mu}) & \ldots & \text{type IIA ,} \\[10pt] G^-(\mu)\,\tilde{G}^-(\bar{\mu}) & \ldots & \text{type IIB ,}\end{array} \right.
\end{align}
where $G^\pm$ ($\tilde{G}^\pm$) are the twisted left-(right-)moving worldsheet supercurrents which are sewed with the Beltrami-differentials $\mu_k$ of $\Sigma_{g,n}$. The supercurrents are part of the $\mathcal{N}=2$ worldsheet superconformal algebra that contains additionally the energy-momentum tensor $T$ ($\tilde{T}$) as well as a $U(1)$ Kac-Moody current $J$ ($\tilde{J}$). More details, including the algebra relations between all operators, are compiled in Appendix~\ref{App:ConformalAlgebra}. 

Furthermore, the insertions $\bar{\phi}_{\inde}$ in (\ref{TwistAmplitude}) are integrated over the full Riemann surface $\Sigma_{g,n}$, while the operators $\hins$ are obtained by folding $\ins$ with the unique holomorphic three-form $\rho$ on the Calabi-Yau space:
\begin{align}
\hins=\oint dz \rho (z)\oint d\bar{z}\tilde{\rho}(\bar{z})\,\ins\,.
\end{align}
The $\hins$ are not integrated over the worldsheet $\Sigma_{g,n}$, but are localised at the positions $u_\ell$ of the $n$ punctures. 

For convenience, we have compiled the charges and (twisted) conformal dimensions of the operators of interest in the following table, distinguishing the type IIA and type IIB setups.
\begin{center}
\begin{tabular}{c||c|c||c|c}\hline
\textbf{operator} & \textbf{charge IIA} & \textbf{twisted dim IIA} & \textbf{charge IIB} & \textbf{twisted dim IIB}\\\hline
&&&&\\[-12pt]
$G^+$ & $(1,0)$ & $(1,0)$ & $(1,0)$ & $(1,0)$\\
&&&&\\[-12pt]
$\tilde{G}^+$ & $(0,1)$ & $(0,2)$ & $(0,1)$ & $(0,1)$\\\hline 
&&&&\\[-12pt]
$G^-$ & $(-1,0)$ & $(2,0)$ & $(-1,0)$ & $(2,0)$\\
&&&&\\[-12pt]
$\tilde{G}^-$ & $(0,-1)$ & $(0,1)$ & $(0,-1)$ & $(0,2)$\\\hline 
&&&&\\[-12pt]
$\ins$ & $(-1,1)$ & $(1,1)$ & $(-1,-1)$ & $(1,1)$\\ 
&&&&\\[-12pt]
$\hins$ & $(2,-2)$ & $(0,0)$ & $(2,2)$ & $(0,0)$\\\hline 
&&&&\\[-12pt]
$\rho$ & $(3,0)$ & $(0,0)$ & $(3,0)$ & $(0,0)$\\
&&&&\\[-12pt]
$\tilde{\rho}$ & $(0,-3)$ & $(0,0)$ & $(0,3)$ & $(0,0)$\\\hline 
\end{tabular}
\end{center}
${}$\\
Notice in particular that the total charges of all insertions in (\ref{TwistAmplitude}) add up to $(-3g+3,\pm 3g\mp 3)$ in the type IIA (type IIB) theory, as is appropriate for a $g$-loop correlator.

The amplitudes $F_{g,n,\bi}$ defined in (\ref{AmpsDefEffAction}) can be computed in a similar manner as the $F_{g,n}$. The only difference is that one of the $A^\mu_\star$ gauge fields is replaced by a different vector multiplet $A^\mu_{\bi}$. At the level of the vertex operators, we simply replace the internal state $\ins$ in (\ref{VertexDef}) and (\ref{VertexInternal}) by another (anti-chiral, chiral) primary ((anti-chiral, anti-chiral) primary) state $\bar{\phi}_{\bi}$ of the type IIA (type IIB) worldsheet theory. Assuming that the gauge fields $A^\mu_\star$ and $A^\mu_{\bi}$ have no contact terms among themselves, we can immediately write the following expression for the amplitude in terms of a twisted worldsheet correlation function
\begin{align}
F_{g,n,\bi}=\int_{\mathcal{M}_{g,n}}\left\langle\prod_{k=1}^{3g-3+n}|\mu_k\cdot G^-|^2\left(\prod_{m=1}^{n-1}\int_{\Sigma_{g,n}} \ins (z_m)\right)\,\left(\int_{\Sigma_{g,n}}\bar{\phi}_{\bi} (z_0)\right)\,\left(\prod_{\ell=1}^n\hins(u_\ell)\right)\right\rangle_{\text{twist}}\,.\label{TwistAmplitudeFi}
\end{align}
Since the charges and (twisted) dimensions of $\bar{\phi}_{\bi}$ are identical to $\ins$, the total charges of all insertions again add up to $(-3g+3,\pm 3g\mp3)$ respectively.
%%%
\subsubsection{Gaugino Amplitudes}
Besides the amplitudes (\ref{TwistAmplitude}) and (\ref{TwistAmplitudeFi}) presented above, the differential equations (\ref{bWEquPredict0}) and (\ref{bWEquPredict}) predicted by supergravity also involve $\Psi^{g,n}_{(\star\star|\bar{I})}$ defined in (\ref{AmpsDefEffAction}). The latter has two insertions of gaugini $\bar{\lambda}^{\star \dot{\alpha}}_a$, whose vertex operators can be obtained from (\ref{VertexDef}) by the action of the supersymmetry generators. Using the same procedure as before, one shows that the amplitude is computed by replacing two of the $\ins$ primary insertions by their superdescendants:
\begin{align}
\Psi^{g,n}_{(\star\star|\bar{I})}&=\int_{\mathcal{M}_{g,n}}\bigg\langle\prod_{k=1}^{3g-3+n}|\mu_k\cdot G^-|^2\left(\prod_{m=1}^{n-2}\int_{\Sigma_{g,n}} \ins \right)\,\left(\int_{\Sigma_{g,n}} \oint G^+\ins \right)\,\left(\int_{\Sigma_{g,n}} \oint \tilde{G}^{\mp}\ins \right)\nonumber\\
&\hspace{2.5cm}\times\left(\int_{\Sigma_{g,n}}\bar{\phi}_{\bar{I}}\right)\,\left(\prod_{\ell=1}^n\hins(u_\ell)\right)\bigg\rangle_{\text{twist}}\,.\label{TwistAmplitudeGaugino}
\end{align}
Again, the total charges of all insertions add up to $(-3g+3,\pm 3g\mp 3)$ in the type IIA (type IIB) theory, respectively.
%%%%%%%%%%%%%%%%%
\subsection{Differential Equations}
%%%%%%%%%%%%%%%%%
\subsubsection{Anti-Holomorphic Derivatives and Operator Insertion}
Having written the relevant couplings in the form of correlation functions of the twisted type II worldsheet theory, we can now derive the stringy analogue of equations (\ref{bWEquPredict0}) and (\ref{bWEquPredict}). In the framework of the twisted worldsheet correlation functions, an anti-holomorphic moduli derivative $\bar{\mathcal{D}}_{\bar{I}}$ corresponds to an insertion of the following operator 
\begin{align}
&\text{type IIA:} &&-\oint G^+ \oint \tilde{G}^-  \bar{\phi}_{\bar{I}}\,,&&\text{charge}(\bar{\phi}_{\bar{I}})=(-1,+1)\,,&&\text{dim}(\bar{\phi}_{\bar{I}})=(1,1)\,,\nonumber\\
&\text{type IIB:} &&-\oint G^+ \oint \tilde{G}^+ \bar{\phi}_{\bar{I}}\,,&&\text{charge}(\bar{\phi}_{\bar{I}})=(-1,-1)\,,&&\text{dim}(\bar{\phi}_{\bar{I}})=(1,1)\,.\label{BRSTdeform}
\end{align}
These types of deformations of the (twisted) worldsheet theory are explained in Appendix~\ref{App:DefsChiralRing}, where also our notation for the chiral ring is presented. Notice that $\phi_I$ is a (chiral,anti-chiral) primary state in type IIA and a (chiral,chiral) primary state in the type IIB theory. Thus, the left hand side of equation (\ref{bWEquPredict}) takes the following form (for convenience, we adopt a streamlined shorthand notation):
\begin{align}
\bar{\mathcal{D}}_{\bar{I}}F_{g,n}=-\int_{\mathcal{M}_{g,n}}\left\langle\prod_{k=1}^{3g-3+n}|\mu_k\cdot G^-|^2\left(\int \ins\right)^n\left(\hins\right)^n\,\left(\int\oint G^+ \oint \tilde{G}^\mp \bar{\phi}_{\bar{I}}\right)\right\rangle_{\text{twist}}\,,
\end{align}
where we are treating the type IIA and type IIB theory simultaneously. Since in the twisted theory $(G^+,\tilde{G}^\mp)$ have dimensions one, we can deform the corresponding contour integrals to encircle different operators in the correlator. We have  
\begin{align}
&\oint G^+ \hins=\oint \tilde{G}^-\hins=0&&\text{in type IIA}\,,\\
&\oint G^+ \hins=\oint \tilde{G}^+\hins=0&&\text{in type IIB}\,,
\end{align}
due to fact that $\hins$ has charge $(+2,\mp2)$. However, there is a non-trivial residue when $G^+$ or $\tilde{G}^\mp$ encircles $\ins$ or one of the operators of the integral measure. Therefore, we find the following contributions
\begin{align}
&\bar{\mathcal{D}}_{\star}F_{g,n}=n\int_{\mathcal{M}_{g,n}}\left\langle\prod_{k=1}^{3g-3+n}|\mu_k\cdot G^-|^2\left(\int \ins\right)^{n-1}\left(\hins\right)^n\,\left(\int \oint G^+ \ins\right)\,\left(\int \oint \tilde{G}^\mp \ins\right)\right\rangle_{\text{twist}}\nonumber\\
&+\int_{\mathcal{M}_{g,n}}\left\langle\sum_{r=1}^{3g-3+n}\prod_{k\neq r}|\mu_k\cdot G^-|^2\,(\mu_r\cdot T)(\bar{\mu}_r\cdot\tilde{G}^{\pm})\,\left(\int \ins\right)^n\left(\hins\right)^{n}\,\left(\int \oint \tilde{G}^\mp \ins\right)\right\rangle_{\text{twist}}\,.\label{StressTensorInsertionStar}
\end{align}
The first line in this relation corresponds to the amplitude $\Psi^{g,n}_{(\star\star|\star)}$. The second line has an insertion of the (left-moving) energy momentum tensor sewed with one of the Beltrami differentials. As we discuss in the next section, such a term can be written as a total derivative  \cite{Bershadsky:1993cx} in the moduli space $\mathcal{M}_{g,n}$ and therefore corresponds to a boundary contribution $\mathcal{C}^{\text{bdy}}_\star$:
\begin{align}
\bar{\mathcal{D}}_{\star}\,F_{g,n}=n\Psi^{g,n}_{(\star\star|\star)}+\mathcal{C}^{\text{bdy}}_{\star}\,.\label{DiffEquNoLimStar}
\end{align}
In a similar fashion as in (\ref{StressTensorInsertionStar}), we can write
\begin{align}
&\bar{\mathcal{D}}_{\bi}F_{g,n}=-n\int_{\mathcal{M}_{g,n}}\left\langle\prod_{k=1}^{3g-3+n}|\mu_k\cdot G^-|^2\,\left(\int \ins\right)^{n-1}\left(\int\oint G^+\oint \tilde{G}^\mp\ins\right)\left(\hins\right)^n\,\int \bar{\phi}_{\bi}\right\rangle_{\text{twist}}\nonumber\\
&-n(n-1)\int_{\mathcal{M}_{g,n}}\left\langle\prod_{k=1}^{3g-3+n}|\mu_k\cdot G^-|^2\,\left(\int \ins\right)^{n-2}\left(\int\oint G^+\ins\right)\left(\int\oint \tilde{G}^{\mp}\ins\right)\left(\hins\right)^n\,\int \phi_{\bi}\right\rangle_{\text{twist}}\nonumber\\
&+\mathcal{C}^{\text{bdy}}_{\bi}\,,\label{StressTensorInsertion}
\end{align}
where for the boundary contribution we write $\mathcal{C}_{\bi}^{\text{bdy}}=\mathcal{C}_{\bi}^{\text{bdy},1}+\mathcal{C}_{\bi}^{\text{bdy},2}$, with
\begin{align}
&\mathcal{C}_{\bi}^{\text{bdy},1}=\int_{\mathcal{M}_{g,n}}\left\langle\sum_{r=1}^{3g-3+n}\prod_{k\neq r}|\mu_k\cdot G^-|^2\,(\mu_r\cdot T)(\bar{\mu}_r\cdot\tilde{T})\,\left(\int \ins\right)^n\left(\hins\right)^n\,\int \bar{\phi}_{\bi}\right\rangle_{\text{twist}}\,,\nonumber\\
&\mathcal{C}_{\bi}^{\text{bdy},2}=-n\int_{\mathcal{M}_{g,n}}\left\langle\sum_{r=1}^{3g-3+n}\prod_{k\neq r}|\mu_k\cdot G^-|^2\,(\mu_r\cdot T)(\bar{\mu}_r\cdot\tilde{G}^{\pm})\,\left(\int \ins\right)^{n-1}\left(\hins\right)^{n}\,\left(\int \oint \tilde{G}^\mp \ins\right)\,\int\phi_{\bi}\right\rangle_{\text{twist}}\nonumber\\
&-n\int_{\mathcal{M}_{g,n}}\left\langle\sum_{r=1}^{3g-3+n}\prod_{k\neq r}|\mu_k\cdot G^-|^2\,(\mu_r\cdot G^-)(\bar{\mu}_r\cdot\tilde{T})\,\left(\int \ins\right)^n\left(\hins\right)^{n}\,\left(\int \oint G^+ \ins\right)\,\int\phi_{\bi}\right\rangle_{\text{twist}}\,.\label{BoundaryDetail}
\end{align}
The first two lines in (\ref{StressTensorInsertion}) can immediately be interpreted as (derivatives of) the amplitudes $F_{g,n,\bi}$ and $\Psi^{g,n}_{(\star\star|\bi)}$ given in (\ref{TwistAmplitudeFi}) and (\ref{TwistAmplitudeGaugino}) respectively.\footnote{Note that the assumption of absence of contact terms allows this re-intepretation.} Concretely, we find
\begin{align}
\bar{\mathcal{D}}_{\bi}\,F_{g,n}=n\,\bar{\mathcal{D}}_\star\,F_{g,n,\bi}-n(n-1)\Psi^{g,n}_{(\star\star|\bi)}+\mathcal{C}^{\text{bdy}}_{\bi}\,.\label{DiffEquNoLim}
\end{align}
The two relations (\ref{DiffEquNoLimStar}) and (\ref{DiffEquNoLim}) are very close the the predicted relations (\ref{bWEquPredict0}) and (\ref{bWEquPredict}) respectively, except for the additional boundary terms $\mathcal{C}^{\text{bdy}}_{\star}$ and $\mathcal{C}^{\text{bdy}}_{\bi}$, which we shall discuss in the following subsection. As already alluded to, these terms receive contributions from the boundaries of $\mathcal{M}_{g,n}$ and encode effects which go beyond the simple on-shell supergravity analysis of Section~\ref{Sect:SugraDiffEqu}.
%%%%%%%%%%%%%%%%%%%%%%%%%%%%%%%%%%%%%%%%
\subsubsection{Boundary Contributions}
The terms $\mathcal{C}^{\text{bdy}}_\star$ and $\mathcal{C}^{\text{bdy}}_{\bi}$ introduced above contain energy momentum tensors sewed with the Beltrami differentials. The latter can be re-written as partial derivatives with respect to the local coordinates of $\mathcal{M}_{g,n}$ and are thus total derivatives. However, $\mathcal{C}_{\bar{I}}^{\text{bdy}}$ are not zero, as one might na\"ively conclude, due to the contributions from boundaries of $\mathcal{M}_{g,n}$. Geometrically, these boundaries correspond to degenerations of $\Sigma_{g,n}$ of which there are three different types:

%\newline
%%
\begin{center}
\begin{tikzpicture}
%dividing geodesic
\node at (-7,0) {$\bullet$ pinching of a dividing geodesic:};
\draw[ultra thick,scale=0.5] (-6,0) to [out=88, in=180] (-4,1.3) to [out=0, in=180] (-2,1) to [out=0, in=180] (0,1.3) to [out=0, in=92] (2,0.1);
\draw[ultra thick,scale=0.5] (-6,0) to [out=272, in=180] (-4,-1.3) to [out=0, in=180] (-2,-1) to [out=0, in=180] (0,-1.3) to [out=0, in=268] (2,-0.1);
\draw[ultra thick, scale=0.5] (1.3,0.5) to [out=315,in=180] (2,0.125) to [out=0,in=180] (3.5,0.125) to [out=0, in=225] (4.2,0.5);
\draw[ultra thick, scale=0.5] (1.3,-0.5) to [out=45,in=180] (2,-0.125) to [out=0,in=180] (3.5,-0.125) to [out=0, in=135] (4.2,-0.5);
\draw[ultra thick,scale=0.5] (-4.9,0.25) to [out=280, in=180] (-4.2,-0.35) to [out=0, in=260] (-3.5,0.25);
\draw[ultra thick,scale=0.5] (-4.75,-0.05) to [out=45, in=135]  (-3.65,-0.05);
\draw[ultra thick,scale=0.5] (-0.7,0.25) to [out=280, in=180] (0,-0.35) to [out=0, in=260] (0.7,0.25) ;
\draw[ultra thick,scale=0.5] (-0.55,-0.05) to [out=45, in=135] (0.55,-0.05);
\draw[ultra thick, scale=0.5] (3.5,0.2) to [out=88,in=180] (5.2,1.3) to [out=0,in=92] (7,0); 
\draw[ultra thick, scale=0.5] (3.5,-0.2) to [out=272,in=180] (5.2,-1.3) to [out=0,in=268] (7,0); 
\draw[ultra thick,scale=0.5,xshift=1cm] (3.5,0.25) to [out=280, in=180] (4.2,-0.35) to [out=0, in=260] (4.9,0.25) ;
\draw[ultra thick,scale=0.5,xshift=1cm] (3.65,-0.05) to [out=45, in=135] (4.75,-0.05);
\node at (-1.9,0.35) {$\bullet$};
\node at (-0.5,-0.3) {$\bullet$};
\node at (4,0) {,};
%\end{tikzpicture}
%\end{center}
%${}$\\
%\begin{center}
%\begin{tikzpicture}
%%%%%%%%%%%%%%%%%%%%%%%%%%%
%handle
\node[yshift=-2cm] at (-7.9,0) {$\bullet$ pinching of a handle:};
\draw[ultra thick,scale=0.5,yshift=-4cm] (-6,0) to [out=88, in=180] (-4,1.3) to [out=0, in=180] (-2,1) to [out=0, in=180] (0,1.3) to [out=0, in=140] (1.5,0.8);
\draw[ultra thick,scale=0.5,yshift=-4cm] (-6,0) to [out=272, in=180] (-4,-1.3) to [out=0, in=180] (-2,-1) to [out=0, in=180] (0,-1.3) to [out=0, in=220] (1.5,-0.8);
\draw[ultra thick, scale=0.5,yshift=-4cm] (1.7,0.6) to [out=300,in=88] (1.875,0) to [out=268,in=60] (1.7,-0.6);
\draw[ultra thick,scale=0.5,yshift=-4cm] (-0.7,0.25) to [out=280, in=180] (0,-0.35) to [out=0, in=260] (0.7,0.25) ;
\draw[ultra thick,scale=0.5,yshift=-4cm] (-0.55,-0.05) to [out=45, in=135] (0.55,-0.05);
\draw[ultra thick,scale=0.5,yshift=-4cm] (-4.9,0.25) to [out=280, in=180] (-4.2,-0.35) to [out=0, in=260] (-3.5,0.25);
\draw[ultra thick,scale=0.5,yshift=-4cm] (-4.75,-0.05) to [out=45, in=135]  (-3.65,-0.05);(7,0); 
\node[yshift=-2cm] at (-2.6,-0.2) {$\bullet$};
\node[yshift=-2cm] at (-0.75,0.15) {$\bullet$};
\draw[ultra thick,scale=0.5,yshift=-4cm] (1.3,0.5) to [out=45, in=180] (2.4,1.4) to [out =0, in=90] (3.4,0) to [out=270,in=0] (2.4,-1.4) to [out=180,in=315] (1.3,-0.5);
\draw[ultra thick,scale=0.5,yshift=-4cm] (1.5,0.3) to [out=45, in=180] (2.4,1.1) to [out =0, in=90] (3.15,0) to [out=270,in=0] (2.4,-1.1) to [out=180,in=315] (1.5,-0.3);
\node at (4,-2) {,};
%%%%%%%%%%%%%%%%%%%%%%%%%%%%
%punctures
\node[yshift=-4cm] at (-7.4,0) {$\bullet$ collision of two punctures:};
\draw[ultra thick,scale=0.5,yshift=-8cm] (-6,0) to [out=88, in=180] (-4,1.3) to [out=0, in=180] (-2,1) to [out=0, in=180] (0,1.3) to [out=0, in=180] (2,1) to [out=0, in=180] (4,1.3) to [out=0, in=92] (6,0);
\draw[ultra thick,scale=0.5,yshift=-8cm] (-6,0) to [out=272, in=180] (-4,-1.3) to [out=0, in=180] (-2,-1) to [out=0, in=180] (0,-1.3) to [out=0, in=180] (2,-1) to [out=0, in=180] (4,-1.3) to [out=0, in=268] (6,0);
\draw[ultra thick,scale=0.5,yshift=-8cm] (-4.9,0.25) to [out=280, in=180] (-4.2,-0.35) to [out=0, in=260] (-3.5,0.25);
\draw[ultra thick,scale=0.5,yshift=-8cm] (-4.75,-0.05) to [out=45, in=135]  (-3.65,-0.05);
\draw[ultra thick,scale=0.5,yshift=-8cm] (3.5,0.25) to [out=280, in=180] (4.2,-0.35) to [out=0, in=260] (4.9,0.25) ;
\draw[ultra thick,scale=0.5,yshift=-8cm] (3.65,-0.05) to [out=45, in=135] (4.75,-0.05);
\draw[ultra thick,scale=0.5,yshift=-8cm] (-0.7,0.25) to [out=280, in=180] (0,-0.35) to [out=0, in=260] (0.7,0.25) ;
\draw[ultra thick,scale=0.5,yshift=-8cm] (-0.55,-0.05) to [out=45, in=135] (0.55,-0.05);
\node[yshift=-4cm] at (-1.1,0.05) {$\bullet$};
\node[yshift=-4cm] at (-0.9,-0.05) {$\bullet$};
\draw[ultra thick, yshift=-4cm,->] (-1.7,0.4) -- (-1.2,0.125);
\draw[ultra thick, yshift=-4cm,->] (-0.3,-0.4) -- (-0.8,-0.125);
\node at (4,-4) {.};
\end{tikzpicture}
\end{center}

\noindent
The first two contributions can be treated in the same manner as in \cite{Bershadsky:1993cx} and are discussed in detail in Appendix~\ref{App:Boundary}. The collision of two punctures is more involved and is proportional to the curvature on the worldsheet, and is not discussed explicitly. However, we remark that its contribution is proportional to $C_{\star\bar{J}\bar{K}}$, and turns out to play no role in our later considerations. Summarising the boundary terms, we find \begin{align}
\mathcal{C}_{\star}^{\text{bdy}}=&\frac{1}{2} {C_{\inde}}^{JK}  \left({\sum_{g',n'}}^{'}\mc{D}_J F_{g',n'} \mc{D}_K F_{g-g',n-n'}+\mc{D}_J \mc{D}_K F_{g-1,n}\right)+\text{(curvature contributions)}\,,\nonumber\\
\mathcal{C}_{\bi }^{\text{bdy}}=&-\frac{1}{2}n\,{C_{\inde\,}}^{JK}\Biggl({\sum_{g',n'}}^{'}\mc{D}_J F_{g',n',\bi } \mc{D}_K F_{g-g',n-n'}+\mc{D}_J \mc{D}_K F_{g-1,n,\bi }\Biggr)\nonumber\\
&\hspace{0.3cm}+\frac{1}{2}{C_{\bi\,}}^{JK}\Biggl({\sum_{g',n'}}^{'} \mc{D}_J F_{g',n'} \mc{D}_KF_{g-g',n-n'}+ \mc{D}_J \mc{D}_K F_{g-1,n}\Biggr)+\text{(curvature contributions)}\,.\nonumber
\end{align}
Here, we have used the shorthand notation
\begin{equation}
 {C_{\bI\,}}^{JM}:= e^{2K}C_{\bI \bJ \bar{M}}G^{\bJ J}G^{\bar{M}M}\,,\label{NotationThree}
\end{equation}
where $K$ is the K\"ahler potential. In addition, and throughout the manuscript, the notation $\Sigma^{'}$ means that we exclude the terms $(0,0),(0,1),(g,n-1)$ and $(g,n)$ from the summation range. Combined with (\ref{DiffEquNoLim}), this gives rise to the following equations which are valid at a generic point in the full string moduli space up to curvature contributions
\begin{align}
\bar{\mathcal{D}}_{\star}\,F_{g,n}&=n\Psi^{g,n}_{(\star\star|\star)}+\frac{1}{2} {C_{\inde}}^{JK}  \left({\sum_{g',n'}}^{'}\mc{D}_J F_{g',n'} \mc{D}_K F_{g-g',n-n'}+\mc{D}_J \mc{D}_K F_{g-1,n}\right)\,,\label{DiffEquNoLimFullStar}\\
\bar{\mathcal{D}}_{\bi}F_{g,n}&=n\,\bar{\mathcal{D}}_\star\,F_{g,n,\bi}-n(n-1)\Psi^{g,n}_{(\star\star|\bi)}-\frac{n}{2}\,{C_{\inde\,}}^{JK}\Biggl({\sum_{g',n'}}^{'}\mc{D}_J F_{g',n',\bi } \mc{D}_K F_{g-g',n-n'}+\mc{D}_J \mc{D}_K F_{g-1,n,\bi }\Biggr)\nonumber\\
&\hspace{0.5cm}+\frac{1}{2}{C_{\bi\,}}^{JK}\Biggl({\sum_{g',n'}}^{'}\mc{D}_J F_{g',n'} \mc{D}_KF_{g-g',n-n'}+ \mc{D}_J \mc{D}_K F_{g-1,n}\Biggr)\,.
\label{DiffEquNoLimFull}
\end{align}
Notice, as a consistency check, that \eqref{DiffEquNoLimFull} reduces to \eqref{DiffEquNoLimFullStar} once $\bi$ is taken to be $\star$.
From the point of view of supergravity, apart from the boundary contribution, the equations (\ref{DiffEquNoLimFullStar}) and (\ref{DiffEquNoLimFull}) agree with the predictions from supersymmetry. In general, `anomalous' contributions like $\mathcal{C}^{\text{bdy}}_{\star}$ or $\mathcal{C}^{\text{bdy}}_{\bi}$ are beyond the simple on-shell analysis performed in Section~\ref{Sect:SugraDiffEqu}, as was pointed out in \cite{Bershadsky:1993cx, Berkovits:1994vy,Berkovits:1998ex,Ooguri:1995cp,Antoniadis:1996qg, Antoniadis:2005sd, Antoniadis:2006mr, Antoniadis:2007cw, Antoniadis:2009nv,Antoniadis:2011hq}.

On the other hand, from the point of view of topological string theory, the derivatives $\bar{\mathcal{D}}_{\bar{I}}$ lead to the insertion of the operators (\ref{BRSTdeform}) into the correlator $F_{g,n}$, which is outside the BRST-cohomology. Therefore, whenever the right hand sides of (\ref{DiffEquNoLimFullStar}) and (\ref{DiffEquNoLimFull}) vanish (up to the anomalous boundary contributions), the $F_{g,n}$ may be interpreted as topological objects. The presence of the $\mathcal{M}_{g,n}$ bulk terms in  $\mathcal{D}_{\inde} F_{g,n,\bi}$ and $\Psi^{g,n}_{(\star\star|\bar{I})}$ indicates that the couplings $F_{g,n}$ generically receive contributions from non-physical states in the topologically twisted theory. In the full string theory, this corresponds to the observation that the $F_{g,n}$ are not BPS-saturated quantities, but also receive contributions from non-BPS states. This can be seen from the formulation of the couplings in (\ref{GeneralDterm}): they are not (BPS-saturated) F-terms, but are rather D-terms, with the $\int d^4\theta\,(\bar{D}\cdot\bar{D})^2$ essentially acting as an integration over the full $\mathbb{R}^{4|8}$ standard superspace.

However, we note that the situation changes for $n=0$. Indeed, in this case, equations (\ref{DiffEquNoLimFullStar}) and (\ref{DiffEquNoLimFull}) reduce to the holomorphic anomaly equation \cite{Bershadsky:1993cx} for the topological amplitudes $F_{g}$, discussed in \cite{Antoniadis:1993ze}. The equation then encodes the stronger property that the couplings $F_g$ are holomorphic functions of the vector multiplet moduli \cite{Antoniadis:1993ze}.
%%%%%%%%%%%%%%%%%

\section{Non-Compact Limit}\label{Sec:LocalLimit}
Although we just explained that the correlation functions (\ref{TwistAmplitude}) (for $n\neq0$) are generically non-topological, we argued in \cite{Antoniadis:2013bja} that the string couplings $F_{g,n}$, with $F_{\star}$ identified with the vector superpartner of the K\"ahler modulus of the dual heterotic K3$\times T^2$ theory, possess numerous properties one would expect from a worldsheet realisation of the genus $g$ free energy of the refined topological string. Most importantly, we showed that when expanded around a particular point in the string moduli space, the $F_{g,n}$ reproduce in the point-particle limit the (perturbative part) of Nekrasov's partition function in the general $\Omega$-background. This result was extended beyond the perturbative level in \cite{Antoniadis:2013mna} and is conceptually a very strong check of our proposal. Given this evidence, it is an interesting and important question to study whether the $F_{g,n}$ can be rendered topological in some appropriate limit in the physical moduli space in which the non-physical states (from the point of view of the twisted theory) decouple in the worldsheet description. This would lead to a vanishing of the bulk contributions in the right hand side of (\ref{DiffEquNoLimFullStar}) and (\ref{DiffEquNoLimFull}). In the framework of supergravity, this corresponds to rendering the effective couplings $F_{g,n}$ in (\ref{AmpsDefEffAction}) holomorphic, such that the respective couplings (\ref{RefinedCouplings}) are BPS-saturated.

The necessity of taking such a limit seems rather natural from the point of view of the gauge theory. Indeed, formulating the $\Omega$-background in four-dimensional space-time requires the presence of a $U(1)$ isometry in the internal manifold. Such isometries are generically not present in compact Calabi-Yau threefolds but may arise  in \emph{non-compact} ones. Therefore, we expect that in an appropriate non-compact limit, the differential equations (\ref{DiffEquNoLimFullStar}) and (\ref{DiffEquNoLimFull}) are simplified due to the presence of additional Ward identities ascribed to the emergent $U(1)$ isometry, such that the $F_{g,n}$ become topological objects. In the following, we analyse necessary conditions for this to occur from the point of view of supergravity and of type II string theory. 
%%%%%%%%%%%%%%%%%%%%%%%%%%%%%%%%%
\subsection{Supergravity Conditions}
The conditions (\ref{RelSUGRA0}) and (\ref{RelSUGRA}) derived in supergravity are solely a consequence of supersymmetry and the structure of the coupling (\ref{GeneralDterm}). In particular, they do not simply encode properties of single tensor components $\mathbb{F}_{g,n}^{I_1\ldots I_{2n-2}}$ as a function of the vector multiplets. Rather, once these relations are translated into the language of scattering amplitudes (\ref{bWEquPredict0}) and (\ref{bWEquPredict}), they relate several different objects, instead of just a single type of them and thus give rise to the bulk terms. In the following, we derive a set of consistent conditions that can be imposed on the component functions $\mathbb{F}_{g,n}$, $\mathbb{F}_{g,n}^{\bi}$, $\mathbb{F}_{g,n}^{\bi\bj}$ etc. directly, such that the resulting equations only involve a single class of objects.
%%%%

More specifically, at the level of the amplitudes we impose that both sides in (\ref{bWEquPredict0}) vanish separately
\begin{align}
\bar{\mathcal{D}}_{\star}\,F_{g,n}=\Psi^{g,n}_{(\star\star|\star)}=0\,,
\label{eqnsst}
\end{align}
and similarly that (\ref{bWEquPredict}) splits into the following two separate equations
\begin{align}
&\bar{\mathcal{D}}_{\bi}\,F_{g,n}=0\,,\label{SplitEqusOrder1}\\
&\bar{\mathcal{D}}_\star\,F_{g,n,\bi}=(n-1)\,\Psi^{g,n}_{(\star\star|\bi)}\,.\label{SplitEqusOrder2}
\end{align}
These condition particularly imply that $F_{g,n}$ are holomorphic functions of the vector multiplet scalars and therefore, the corresponding effective action term (\ref{RefinedCouplings}) is BPS-saturated. The equations (\ref{SplitEqusOrder1}) and (\ref{SplitEqusOrder2}) immediately translate to the level of the component couplings as
\begin{align}
&\bar{\mathcal{D}}_{\star}\,T_{g,n}^{(1)}=0\,,&&\text{and} &&T_{g,n}^{(2)}=0\,,\\
&\bar{\mathcal{D}}_{\bi}\,T_{g,n}^{(1)}=0\,,&&\text{and} &&\bar{\mathcal{D}}_\star\,T^{(1)}_{g,n;\bi}=T_{g,n;\bi}^{(2)}\,,
\end{align}
which in turn, respectively, can be written as follows: 
\begin{align}
&\bar{\mathcal{D}}_\star^3\,\mathbb{F}_{g,n}=0\,,\nonumber\\
&\bar{\mathcal{D}}_\star^2\bar{\mathcal{D}}_{\bi}\,\mathbb{F}_{g,n}=0\,,&&\text{and} &&\bar{\mathcal{D}}_\star(\bar{\mathcal{D}}_\star\bar{\mathcal{D}}_{\bi}\mathbb{F}_{g,n}+\bar{\mathcal{D}}_\star^2\mathbb{F}_{g,n;\bi})=\bar{\mathcal{D}}_\star^3\mathbb{F}_{g,n;\bi}\,.
\end{align}
Notice that the latter are equivalent to imposing
\begin{align}
\bar{\mathcal{D}}_\star^2\bar{\mathcal{D}}_{\bar{I}}\,\mathbb{F}_{g,n}=0 \,,\label{ConditionsImpose}
\end{align}
on the effective coupling function $\mathbb{F}_{g,n}$. This constitutes an explicit condition on the moduli dependence going beyond the constraints coming from supersymmetry or T-duality. We expect that (\ref{ConditionsImpose}) may be interpreted as a direct consequence of the $U(1)$ isometry so that in the full quantum theory, (\ref{ConditionsImpose}) corresponds to a Ward identity constraining the moduli dependence of the effective action couplings. Naturally, the specific action on the individual fields and its geometric interpretation in terms of the Calabi-Yau manifold heavily depend on the specific model under consideration and is not analysed here.

As a final remark, we note that (\ref{ConditionsImpose}) is only a condition on the anti-holomorphic moduli dependence of $\mathbb{F}_{g,n}$. The fact that we are treating holomorphic and anti-holomorphic vector multiplet moduli differently is very reminiscent of the holomorphic limit introduced in \cite{Bershadsky:1993cx} (see also \cite{Klemm:2005pd} for an application) which is relevant in extracting topological information from the amplitudes $F_{g,n=0}$.

%%%%%%%%
\subsection{Type II Conditions}
In the previous section, we derived conditions on the moduli dependence of the $\mathbb{F}_{g,n}$ at the level of the effective supergravity action. At the full string theory level, we expect the consequences of (\ref{ConditionsImpose}) to be more involved: the conditions \eqref{eqnsst} -- \eqref{SplitEqusOrder2} still eliminate the bulk terms on the right hand side of eqs.\eqref{DiffEquNoLimFullStar} and \eqref{DiffEquNoLimFull} and therefore ensure that the $F_{g,n}$ defined in (\ref{TwistAmplitude}) are topological. However, we expect \eqref{eqnsst} -- \eqref{SplitEqusOrder2} to be modified by boundary contributions in a non-trivial fashion. While the type II setup provides a well-posed framework to study these modifications, it is difficult to analyse them in full generality, \emph{i.e.} without considering a specific limit for a particular Calabi-Yau compactification. 

A more basic question is whether (\ref{DiffEquNoLimFull}) can take the form of a recursive equation in the decompactification limit, such that the right hand side only contains $F_{g'',n''}$ with $(g'',n'')<(g,n)$. Indeed, such an equation was postulated in \cite{Huang:2010kf, Huang:2011qx} as the definition of the free energy of the refined topological string on local/non-compact Calabi-Yau manifolds and was termed generalised holomorphic anomaly equation. In fact, the right hand side of the latter is very similar to the second line of (\ref{DiffEquNoLimFull}). However, if we indeed assume that \eqref{SplitEqusOrder1} is modified in the following way in the full string-theory setting
\begin{align}
\mathcal{D}_{\bi}F_{g,n}\big|_{\text{lim}}&=\frac{1}{2}{C_{\bi\,}}^{JK}\Biggl({\sum_{g',n'}}^{'}\mc{D}_J F_{g',n'} \mc{D}_KF_{g-g',n-n'}+ \mc{D}_J \mc{D}_K F_{g-1,n}\Biggr)\,,\label{EqLocalLim}
\end{align}
we simultaneously have to require the following modification for \eqref{SplitEqusOrder2} (up to curvature contributions)
\begin{align}
\bar{\mathcal{D}}_\star\,F_{g,n,\bi}\big|_{\text{lim}}&=(n-1)\Psi^{g,n}_{(\star\star|\bi)}+\frac{1}{2}\,{C_{\inde\,}}^{JK}\Biggl({\sum_{g',n'}}^{'}\mc{D}_J F_{g',n',\bi } \mc{D}_K F_{g-g',n-n'}+\mc{D}_J \mc{D}_K F_{g-1,n,\bi }\Biggr)\bigg|_{\text{lim}}\,.\label{ConsistEqLocalLim}
\end{align}
From the perspective of the effective string couplings, (\ref{ConsistEqLocalLim}) plays the role of a consistency condition which supplements (\ref{EqLocalLim}) and is imposed by supersymmetry. We note again that (\ref{EqLocalLim}) and (\ref{ConsistEqLocalLim}) only contain physical objects, (\emph{i.e.} string theory scattering amplitudes), and checking them in a specific realisation is therefore a well-posed problem. Indeed, in the following section we reproduce (\ref{EqLocalLim}) in a specific decompactification limit of the dual heterotic setup on $K3\times T^2$.

%%%%%%%%%%%%%%%%%%%%%%%%%%%%%%%%%%%%%%%%
\section{Heterotic Realisation}\label{Sect:HetRealisation}
The results of the previous sections lend further support to our proposal that the $F_{g,n}$ studied in \cite{Antoniadis:2013bja} can furnish a worldsheet description of the refined topological string for specific choices of the internal manifold or in suitable decompactification limits. The crucial  property for these couplings is that in the point particle limit, the $F_{g,n}$ reproduce Nekrasov's gauge theory partition function on the full $\Omega$-background, with both deformation parameters being non-trivial. In \cite{Antoniadis:2013bja,Antoniadis:2013mna}, working in the dual heterotic theory on $K3\times T^2$, we showed that the $F_{g,n}$ involving insertions of field-strengths in the vector multiplet of the $T^2$ K\"ahler modulus correctly reproduce the perturbative and non-perturbative parts of the Nekrasov partition function when expanded around a particular point in the string moduli space. We denote these couplings by $F_{g,n}^{\bar{T}}$ in the remainder of the section. 

An  important check of the approach described in the previous sections concerns the differential equations satisfied by the realisation of the couplings $F_{g,n}$ in the dual heterotic framework on $K3\times T^2$, since their explicit expression is known by a direct one-loop computation at the full string level \cite{Antoniadis:2013bja}. We show in the following that, in the large $T^2$ volume limit, the equations satisfied by $F_{g,n}^{\bar T}$ precisely match with the weak coupling limit of \eqref{EqLocalLim}.

%%%%%%%%%%%%%%%%%%%%%%
\subsection{Heterotic One-Loop Couplings}\label{Sect:HetCouplings}

As discussed in Section~\ref{Sec:Sugra}, the couplings $F_{g,n}$ at the component level contain terms involving two self-dual Riemann tensors $R_{(-)}$, $(2g-2)$ self-dual graviphoton field strength tensors $F^G_{(-)}$ and $2n$ anti-self-dual vector multiplet field strength tensors $F_{(+)}^\star$. In the heterotic compactification, the vector multiplet moduli space is given by the product of coset manifolds
\begin{align}
&\mathcal{M}^{\text{het}}_{\text{vec}}=\frac{SU(1,1)}{U(1)}\times \frac{O(2,10)}{O(2)\times O(10)}\in (S;T,U,W^a)\,,&&a=1,\ldots,8\,,\label{ModuliSpace}
\end{align}
which we parametrise by complex variables $S,T,U$ and $W^a$. Physically, they correspond to the heterotic dilaton, the K\"ahler, complex structure moduli of $T^2$ and Wilson lines respectively. In order to compute the $F_{g,n}$ at the one-loop level in heterotic string theory, the relevant piece of information is the vertex operator of the vector superparnters $F_{(+)}^\star$ of these moduli. For the moduli in the coset $\frac{O(2,10)}{O(2)\times O(10)}$, they take the form
\begin{align}
V^\star (p,\eta;z)=\eta_\mu\left[\partial Z^\mu -i(p\cdot \chi)\chi^\mu\right](z)\,\bar{\mathcal{J}}^\star(\bar{z})\,e^{ip\cdot Z}\,.\label{VertStarGauge}
\end{align}
In this work we adopt a notation similar to \cite{Antoniadis:2013mna} and consider an orbifold representation of $K3$, such that $(Z_\mu,Z_3,Z_{4,5})$ denote the complex coordinates of space-time, the $T^2$-torus and $K3$ respectively, with $(\chi^\mu,\chi^3,\chi^{4,5})$ being their fermionic superpartners. Furthermore, $\eta^\mu$ and $p^\mu$ denote the polarisation and momentum of the gauge field with $p\cdot \eta=0$, while $z$ is the worldsheet position. The relevant quantity distinguishing different multiplets is the right-moving current $\bar{\mathcal{J}}^\star$. It can either be a bosonic current of $T^2$, \emph{i.e.} $\bar{\partial} Z_3$ or $\bar{\partial} \bar{Z}_3$, or a current of the $E_8$ gauge group. For later convenience, we organise the latter in a complex basis  $(\bar{J}^m,\bar{J}^{m\dagger})$, with $m=1,\ldots, 4$. In \cite{Antoniadis:2013bja}, we chose $\bar{\mathcal{J}^\star}=\bar{\partial} Z_3$, giving $V^\star$ the interpretation of the vector 
superpartner of the $\bar{T}$-modulus of $T^2$, as already  mentioned above.

To explicitly compute the couplings $F_{g,n}$ at the one-loop level, we can follow the same strategy as in \cite{Antoniadis:2013bja} and introduce the generating functional
\begin{align}
\mathcal{F}(\epsilon_-,\epsilon_+)=\sum_{g,n} \frac{\epsilon_+^{2n}}{(n!)^2}\,\frac{\epsilon_-^{2g-2}}{((g-1)!)^2}\,F_{g,n}\,, \label{GenFunc}
\end{align}
which can be computed as the partition function of a deformed worldsheet sigma model whose action is
\begin{align}
S=S_{\text{free}}&-\int{d^2 z}\left[\epsilon_-\,\partial Z_3\left(Z^1 \bar\partial Z^2 + 
\bar{Z}^2\bar\partial\bar{Z}^1\right)+\epsilon_{+}\left(Z^1 \partial \bar{Z}^2 + Z^2\partial\bar{Z}^1+\chi^4\chi^5-\bar{\chi}^4\bar{\chi}^5\right)\bar{\mathcal{J}}^\star\right]\,.
	\label{DefAction}
\end{align}
It is important to notice that, provided
\begin{align}
\langle \bar{\mathcal{J}}^\star (\bar{z})\,\bar{\mathcal{J}}^\star (\bar{w})  \rangle=0\,,&&\text{and}&&\langle \bar{\mathcal{J}}^\star (\bar{z})\,\partial Z_3 (\bar{w})  \rangle=0\,,\label{CondExact}
\end{align}
this action is exact, wheareas otherwise it receives additional $\alpha'$-corrections. Specifically, since the internal currents are formulated in a complex basis such that
\begin{align}
&\langle {J}^k(\bar{z}){J}^m(0)\rangle= 0\,,&&\langle{J}^k(\bar{z}) {J}^m(0)^\dagger\rangle= \frac{\delta^{km}}{\bar{z}^2}\,,
\end{align}
the deformation (\ref{DefAction}) is not only exact if $F^\star$ is chosen to be the vector superpartner of the $\bar{T}$-modulus, but also if it is identified with any one of the gauge fields of the $E_8$ group.\footnote{However, when choosing $V^\star$ to be the vector partner of the $\bar{U}$-modulus (with $\bar{\mathcal{J}}^\star$ being identified with $\bar{\partial} \bar{Z_3}$), there are additional corrections to (\ref{DefAction}) as well as (\ref{HetFormFgn}) below, since the second relation in (\ref{CondExact}) is not satisfied.} In these cases, the generating functional (\ref{GenFunc}) can be computed exactly and, after expanding in powers of $\epsilon_{\pm}$, we find
\begin{align}
F_{g,n}=\int_{\mathfrak{F}}\frac{d^2\tau}{\tau_2^2}~G_{g,n}(\tau,\bar\tau)\,\tau_2^{2g+2n-1} \sum_{m_i,n^i,b^a\in\Gamma^{2,10}} \left(\frac{P_L}{\xi}\right)^{2g-2}\left(\frac{P_R^\star}{\xi}\right)^{2n}\,q^{\frac{1}{2}|P_L^2|}\,\bar{q}^{\frac{1}{2}\vec{P}_R\cdot \vec{P}_R}\,,\label{HetFormFgn}
\end{align}
where the integral is over the fundamental domain $\mathfrak{F}$ of $\textrm{SL}(2,\mb Z)$, parametrised by $\tau=\tau_1+i\tau_2$ living in the upper half-plane $\mathcal{H}_+$, and we also use $q=e^{2\pi i \tau}$. Furthermore, $G_{g,n}(\tau,\bar{\tau})$ is a non-holomorphic modular form which was computed explicitly in \cite{Antoniadis:2013bja} and the summation in (\ref{HetFormFgn}) is over the $\Gamma^{2,10}$ self-dual lattice parametrised by momenta $(P_L,\bar{P}_L;\vec{P}_R)$ (for our conventions concerning the latter, we refer the reader to Appendix~\ref{App:Lattice}). The shorthand notation $\xi$ is introduced in (\ref{XiDef}). The momentum insertion $P^\star_R$ in (\ref{HetFormFgn}) denotes a particular component of $\vec{P}_R$. It reflects the choice of $V^\star$ since it is the zero mode of the current $\bar{\mathcal{J}}^\star$ in (\ref{VertStarGauge}). 
%%%%%%%%%%%
\subsection{Differential Equations}
We now use the explicit expression for the $F_{g,n}$ to test some of the ideas advocated in the previous section. In particular, we study decompactification limits of the $K3\times T^2$ internal geometry and compare them to (\ref{EqLocalLim}) in the weak coupling regime. For convenience, we consider the limit corresponding to large torus volume, and work with $F_{g,n}^{\bar{T}}$, \emph{i.e.} $F^\star$ identified with the field strength of the $\bar{T}$-vector. The results can be straightforwardly generalised to the other possible insertions without altering the main conclusions.
%%%%%%%%%%%
\subsubsection{\texorpdfstring{One-Loop Differential Equations for $F_{g,n}^{\bar{T}}$}{One-loop Differential Equations for Fg,n}}\label{Sect:OneLoopTEq}
When choosing $A^\star_\mu$ to be the vector superpartner of the $\bar{T}$-modulus of $T^2$ as in \cite{Antoniadis:2013bja}, we have $P_R^\star=P_R$ in (\ref{HetFormFgn}). We are interested in studying these $F_{g,n}^{\bar{T}}$ in the decompactification limit $\frac{T-\bar{T}}{2i}=T_2\to \infty$. Using the explicit representation of the lattice partition function, we can write\footnote{In the following, we discuss the heterotic equivalent of eq.~(\ref{DiffEquNoLimFull}). While a similar analysis can be made for (\ref{DiffEquNoLimFullStar}), we do not discuss it in this paper.} 
\begin{align}
\mathcal{D}_{\bi}F_{g,n}^{\bar{T}}\big|_{T_2\to\infty}=\frac{i}{2\pi}\,e^{2\tilde{K}}\,C_{\bi\bj\bar S}\,G^{\bj j}\int_{\mathfrak{F}}\frac{d^2\tau}{\tau_2^2}~G_{g,n}(\tau,\bar\tau)\,\tau_2^{2g+2n-1} \partial_\tau\sum_{m_i,n^i,b^a}\left(\tau_2^2\,\mathcal{D}_j K_{g-1,n}\right)\bigg|_{T_2\to\infty}\,,\nonumber
\end{align}
where $\tilde K$ is the K\"ahler potential stripped off its dilaton dependence:
\begin{align}
	\tilde K =-\log [(T-\bar T)(U-\bar U)-(W-\bar W)^2] \,.
\end{align}
Integrating by parts, and using the modular invariance of the integrand, we find
\begin{align}\label{GenCovarEqHet}
\mathcal{D}_{\bi}F_{g,n}^{\bar{T}}\big|_{T_2\to\infty}=\frac{1}{2\pi i}\,e^{2\tilde{K}}\,C_{\bi\bj\bar S}\,G^{\bj j}\,\mathcal{D}_j \tilde{F}^{\bar{T}}_{g-1,n}\big|_{T_2\to\infty}\,,
\end{align}
where
\begin{align}
\tilde{F}^{\bar{T}}_{g-1,n}\equiv \int_{\mathfrak{F}}\frac{d^2\tau}{\tau_2^2}\,(\partial_\tau G_{g,n})\,\tau_2^{2g+2n-1}\sum_{m_i,n^i,b^a}\left(\frac{P_L}{\xi}\right)^{2g-4}\left(\frac{P_R}{\xi}\right)^{2n}\hat\Gamma^{(2,10)}\,.
\end{align}
The limit $T_2\to \infty$ simplifies (\ref{GenCovarEqHet}) by constraining the type of states which can propagate on the worldsheet. Indeed, in the next step, we perform the change of variable $\tau_2\to \tau_2\, T_2$. Notice that the worldsheet torus degenerates in this limit, such that the contribution of higher stringy modes in (\ref{GenCovarEqHet}) is suppressed. More specifically, we can separate the integral over the fundamental domain $\mathfrak{F}$ into orbits of the modular group, a procedure which is known as the unfolding \cite{Unfolding} (see \cite{DevelopUnfolding} and most recently \cite{NewUnfolding} for further developments of these techniques). To be precise, we unfold against the $(2,10)$ self-dual lattice. In the limit of large $T_2$, most contributions are exponentially suppressed, except for the so-called degenerate orbit, which can be written in the form of an integral over the semi infinite strip
\begin{align}
\mathcal{S}=\{\tau\in\mathcal{H}_+:\tau_2>0, |\tau_1|\leq 1/2\}
\end{align} 
with vanishing winding numbers $n^1=n^2=0$ in the lattice momenta. More explicitly, one finds the following recursive relation for the non-holomorphic modular functions $G_{g,n}(\tau,\bar{\tau})$:
\begin{align}
\partial_\tau G_{g,n}\big|_{T_2\to\infty} = \frac{\pi}{2i\tau_2^2}\,G_{g-1,n}\big|_{T_2\to\infty}\,.\label{RecursGgn}
\end{align}
Remarkably, we find the same type of recursion relation as in the case of $G_{g,n=0}$ derived in \cite{Antoniadis:1995zn}. In both cases, this recursive structure stems from a dependence of the modular form on the (extended) second Eisenstein series $\hat{E}_2=E_2-\frac{\pi}{3\tau_2}$, which may be regarded as a form of modular stringy regularisation arising when operators collide on the worldsheet torus. In \cite{Antoniadis:1995zn}, it was shown to be responsible for the \emph{holomorphic anomaly} and is an inherently stringy effect (see also \cite{Haghighat:2013gba,Hohenegger:2013ala,Hohenegger:2015cba} for some recent related work).

Equation (\ref{RecursGgn}) turns (\ref{GenCovarEqHet}) into a recursion in $g$ in the large volume limit, since
\begin{align}
\tilde{F}^{\bar{T}}_{g-1,n}\big|_{T_2\to\infty} = -\frac{i\pi}{2}\,F^{\bar{T}}_{g-1,n}\big|_{T_2\to\infty}
\end{align}
and consequently, we obtain
\begin{align}
\mathcal{D}_{\bi}F_{g,n}^{\bar{T}}\big|_{T_2\to\infty}=\frac{1}{2\pi i}\,e^{2\tilde{K}}\,C_{\bi\bj\bar S}\,G^{\bj j}\,\mathcal{D}_j F_{g-1,n}^{\bar{T}}\big|_{T_2\to\infty}\,.\label{LimEquHet}
\end{align}
Notice that while it is generically non-trivial, from the point of view of the heterotic theory,  to distinguish a bulk term of the type II twisted worldsheet theory from a boundary one, in the case of (\ref{LimEquHet}), the right hand side is a pure boundary contribution. Indeed, due to the recursive structure, the amplitudes on the right hand side are dual to type II correlators of genus $g-1$, which are boundary contributions at genus $g$. Therefore, at least at weak coupling, the $F_{g,n}^{\bar{T}}$ are rendered topological in the decompactification limit $T_2\to \infty$. 
%%%%%
\subsubsection{Weak Coupling Limit}
In order to compare (\ref{LimEquHet}) with the explicit form of (\ref{EqLocalLim}), we first derive the weak coupling limit of the latter equation. To this end, we recall that the two derivative action for the vector multiplets is completely determined in terms of the holomorphic prepotential $F(X)$. In particular, the K\"ahler potential is given by
\begin{align}
	K = - \log[ i(\bar{\varphi}^I F_I-\varphi^I \bar{F}_I)]\,,
\end{align}
where the K\"ahler metric takes the form $G_{I\bar{J}}=\partial_I\partial_{\bar{J}}K$. Explicitly, the classical piece of $F$ is given by
\begin{align}
&F=\frac{S(TU-\frac{1}{2}W^2)}{\varphi^0}\,\label{Potentials}\,,
\end{align}
where $\varphi^0$ is a compensating field. 

Since we are interested in the weak coupling limit $(S-\bar{S})\to\infty$ of (\ref{EqLocalLim}), we consider the following heterotic perturbative expansion:
\begin{align}
 F_{g,n}^{\bar{T}}=\alpha_{g,n}(S-\bar S)+\beta_{g,n}+\mathcal{O}((S-\bar S)^{-1})\,.\label{ExpansionDil}
\end{align}
It was shown in \cite{Antoniadis:1995zn} that $F_{g\geq2,n=0}$ (\emph{i.e.} in the absence of the $F^\star$ insertions) is independent of the heterotic dilaton $S$ at weak-coupling and starts receiving contributions at one-loop, while $F_{g=1,n=0}$ receives a constant tree-level contribution:
\begin{align}
\alpha_{g\geq 2,0}=0\,,&&\alpha_{1,0}=-i\pi\,.
\end{align} 
As before, we analyse the couplings $F_{g,n}^{\bar{T}}$ with $n\neq 0$ in the limit $\frac{T-\bar{T}}{2i}=T_2\to \infty$. Plugging this result into \eqref{EqLocalLim} for $\bi=\bar S$, we obtain
\begin{equation}
\mathcal{D}_{\bar S}F_{0,2}^{\bar T}\big|_{T_2\to\infty}=\mc{O}((S-\bar S)^0)\,.\label{EquIteration2}
\end{equation}
We have replaced the decompactification limit by $T_2\to\infty$ and $\mathcal{D}_i=(\partial_i-w\partial_i K)$ is the K\"ahler covariant derivative acting on a function of weight $w$. From (\ref{EquIteration2}), it follows immediately that $\alpha_{0,2}\big|_{T_2\to\infty}=0$.

We can iterate this analysis in two ways. Firstly, by inserting this result into (\ref{EqLocalLim}) for higher values of $n$, we find that $\alpha_{0,n\geq1}\big|_{T_2\to\infty}=0$. Secondly, we can also extend this iteration for higher values of $g$. To this end, we consider (\ref{EqLocalLim}) for $g=n=1$ and $\bi=\bar{S}$:
\begin{align}\label{Logterms}
 \mathcal{D}_{\bar S}F_{1,1}^{\bar T}\big|_{T_2\to\infty}&=\mathcal{O}((S-\bar{S})^0)\,,			     
\end{align}
so that $\alpha_{1,1}\big|_{T_2\to\infty}=0$. Therefore, by induction, we obtain
\begin{equation}\label{ValueAlpha}
\alpha_{g,n}\big|_{T_2\to\infty}=-i\pi\delta_{g,1}\delta_{n,0}\,,
\end{equation}
\emph{i.e.} in the large $T_2$ limit, only the prepotential and $F_{1,0}$ receive a tree-level contribution.

This expression allows us to formulate the weak-coupling limit of \eqref{EqLocalLim} for $\bi\neq \bar{S}$. Notice that to leading order, given the form of the prepotential \eqref{Potentials}, one of the indices of the Yukawa couplings $C_{\bi\bj\bar k}$ (and therefore also one of the $\mathcal{D}_j$-derivatives) on the right-hand side of \eqref{EqLocalLim}, must correspond to $\bar S$. However, in the weak-coupling limit, only $F_{1,0}$ depends on $\bar{S}$ such that, for $g+n\geq2$, we find
\begin{equation}
 \mathcal{D}_{\bi} F_{g,n}^{\bar T}\big|_{T_2\to\infty}=\frac{1}{2\pi i}\,C_{\bi\bj\bar S}\,e^{2\tilde K}G^{\bj j}\,\mathcal{ D}_j\,F_{g-1,n}^{\bar T}\big|_{T_2\to\infty}\,.
 \label{WeakCouplingHet}
\end{equation}
This matches precisely eq.~(\ref{LimEquHet}) derived in the previous section for the couplings $F_{g,n}^{\bar{T}}$ and provides a non-trivial check for our approach, as discussed in Section~\ref{Sec:LocalLimit}. In particular, it provides additional evidence that the physical realisation we proposed in \cite{Antoniadis:2013bja} for the $F_{g,n}$ provides a viable candidate for a worldsheet realisation of the refined topological string.

%%%%%%%%%%%
\subsection{Nekrasov Partition Function and Boundary Conditions}

In \cite{Huang:2010kf, Huang:2011qx}, the Nekrasov partition function was interpreted as a boundary condition for  eq.~(\ref{EqLocalLim}) and is thus part of the definition of the refined topological string. It has also been shown that consistent solutions to the coupled system exist. From the approach advocated in the previous sections, which takes the effective couplings $F_{g,n}$ as its starting point, it is interesting to study whether the example $F_{g,n}^{\bar{T}}$ is the only class of couplings which captures the Nekrasov partition function in the point particle limit. In particular, since the equations (\ref{DiffEquNoLimFull}) and (\ref{EqLocalLim}) can be formulated for a generic multiplet $X^\star$, \emph{i.e.} covariantly with respect to the T-duality group of the string compactification, it would seem surprising if  the boundary conditions were to break covariance by singling out one specific modulus. 

Concretely, in Section~\ref{Sect:HetCouplings}, we discussed a whole family of different heterotic couplings (\ref{HetFormFgn})  whose one-loop representation (to leading order in $\alpha'$) only differs by the insertions of the right-moving momenta $P_R^\star$. However, in \cite{Antoniadis:2013bja}, it was argued that $P_R^\star=P_R$ was crucial for reproducing the Nekrasov partition function in the field theory limit, when expanding around a so-called Wilson line enhancement point in the heterotic string moduli space. It is therefore very interesting to see whether also the more general couplings (\ref{HetFormFgn}) yield the gauge theory partition function at an appropriate point in the moduli space. This section is devoted to addressing this question.

For simplicity, we restrict the presentation to perturbative corrections.\footnote{Based on T-duality, we expect that the results in \cite{Antoniadis:2013mna} hold  for all choices of $X^\star\in\frac{O(2,10)}{O(2)\times O(10)}$ in (\ref{ModuliSpace}).} In this case, the first step to recovering the Nekrasov partition function is to find a point of enhanced gauge symmetry in the string moduli space at which both $P_L$ and $P_R^\star$ vanish simultaneously. The rate at which these momenta go to zero is proportional to the mass of the BPS states which are responsible for the gauge symmetry enhancement. These vector multiplet states generically depend on $P_R^\star$, \emph{i.e.} the type of insertion $A^\star_\mu$ used for the coupling $F_{g,n}$, and we focus on the case of a pure $SU(2)$ gauge theory.

Using the explicit expressions for the lattice momenta given in Appendix~\ref{App:Lattice}, it is straightforward to analyse the various possibilities for $P^\star_R$, as already explained in Section~\ref{Sect:HetCouplings}
\begin{itemize}
\item $P_R^\star=P_R$:\\
The case $A^\star_\mu=A^{\bar{T}}_\mu$ (leading to $P_R^\star=P_R$) was already discussed at length in \cite{Antoniadis:2013bja}. Indeed, it was found that for
\begin{align}\label{EnhancementPointT}
	(V^{(0)}_{1})^{a}=(V_{2}^{(0)})^{a}=(\tfrac{1}{2},\tfrac{1}{2},v^3,\ldots,v^8) ~~,~~ 
\end{align}
with generic $v^{3,\ldots,8}$, the states characterised by
\begin{align}	
&(m_i,n^i) =0\,,&&{b}^{a}=\pm( 1,- 1,0,\ldots,0) \,,\label{StatebarT}
\end{align} 
become massless and, furthermore, both the left- and right- moving momenta vanish at the same rate:
\begin{align}
	P_L=P_R= \frac{V_2^a b^a-U\, V_1^a b^a}{\sqrt{(T-\bar{T})(U-\bar{U})-\tfrac{1}{2}(\vec{W}-\vec{\bar{W}})^2}} ~\longrightarrow~ 0~.
\end{align}
\item $P_R^\star=\bar{P}_R$:\\
If $A^\star_\mu$ is identified with the vector superpartner of the $\bar{U}$-modulus (in which case $P_R^\star=\bar{P}_R$), at a similar enhancement point like (\ref{EnhancementPointT}) the states (\ref{StatebarT}) become massless and
\begin{align}
P_L=\bar{P}_R\longrightarrow 0\,.
\end{align}
\item $P_R^\star=P^a_R$:\\
Finally, if $A^\star_\mu$ is identified with one of the $E_8$ field strength tensors (such that $P_R^\star=P_R^a$) we can consider
\begin{align}\label{EnhancementPointW}
&(V_1^{(0)})^{a}=(\bar T-\bar U)(1,\ldots,1)\,,&&(V_2^{(0)})^{a}=\textrm{fixed} \,.
\end{align} 
In the limit $T=U$, the (winding) states
\begin{align}\label{WindingStates}
	(m_1,n^1)^\star =(\pm1,\mp1) ~~,~~ (m_2,n^2)^\star=(0,0) ~~,~~ {b}^{a\star}=0  ~,
\end{align}
become massless and
\begin{equation}
	P_L\propto P_R^a\propto \bar T-\bar U\longrightarrow0\,.
\end{equation}
\end{itemize}
In all three cases, by analysing the contribution of the massless states to the worldsheet integral, we precisely reproduce Nekrasov's partition function in the field theory limit. The analysis precisely parallels the one given in \cite{Antoniadis:2013bja} and is not reproduced here. The fact that different choices of the gauge field $A_\mu^\star$ reproduce the Nekrasov partition function in a suitable field theory limit is consistent with the fact that the corresponding amplitudes $F_{g,n}$ are related to one another by T-duality transformations, which are unbroken by the boundary conditions imposed in the point-particle limit.
%%%%%%%%%%%
\section{Interpretation and Conclusions}\label{Sec:Conclusions}
In this paper, we have discussed the class of superspace couplings (\ref{GeneralDterm}) in the $\mathcal{N}=2$ supergravity action. We have analysed consistency conditions between its various component terms that are imposed by supersymmetry. These do not simply constrain the moduli dependence of a single component coupling (\emph{e.g.} holomorphicity as in the case of $n=0$, see \cite{Antoniadis:1993ze}), but rather relate different component terms with one another. These relations were formulated as first order differential equations, \emph{e.g.} (\ref{bWEquPredict0}) and (\ref{bWEquPredict}).

Based on the evidence in support of our proposal \cite{Antoniadis:2013bja} for the $F_{g,n}$ as candidates for the refinement of the topological string, following  \cite{Antoniadis:2010iq}, we derived all couplings~(\ref{AmpsDefEffAction}) as higher loop scattering amplitudes in the framework of type II string theory  on a (compact) Calabi-Yau manifold. These string effective couplings were shown to satisfy (\ref{bWEquPredict0}) and (\ref{bWEquPredict}) up to additional terms which arose as boundary contributions of the moduli space of the genus $g$ worldsheet with $n$ punctures. The latter play a similar role as the \emph{holomorphic anomaly} found in \cite{Antoniadis:1993ze} in the case of $n=0$. The resulting equations (\ref{DiffEquNoLimFullStar}) and (\ref{DiffEquNoLimFull}) are solely a consequence of the $\mathcal{N}=(2,2)$ worldsheet supersymmetry and hold at a generic point in the string moduli space. Provided certain well-defined conditions are met, these equations reduce to a form involving only 
one type of component couplings and exhibit a recursive structure in both $g$ and $n$. The resulting equation (\ref{EqLocalLim}) is structurally similar to the generalised holomorphic anomaly equation  proposed in \cite{Huang:2010kf, Huang:2011qx} as a definition for the free energy of the refined topological string on local/non-compact Calabi-Yau manifolds.

These results support our  proposal \cite{Antoniadis:2013bja,Antoniadis:2013mna} for the couplings $F_{g,n}$ as a worldsheet definition of the refined topological string. The present work further analyses the necessary conditions for the validity of our proposal. At a generic point in the moduli space of a (compact) Calabi-Yau manifold, the couplings $F_{g,n}$  are not BPS-saturated and their (twisted) worldsheet representation (\ref{TwistAmplitude}) is not topological. This manifests itself in the fact that the $F_{g,n}$ are related to different classes of couplings. We expect that the $U(1)$ isometry, recovered at certain regions in the boundary of moduli space, is responsible for a simplification of these equations (see \emph{e.g.} (\ref{EqLocalLim})) that is appropriate for a topological object. We have provided the well-posed necessary and sufficient conditions (\ref{ConsistEqLocalLim}) (formulated in terms of physical quantities only) for this modification to happen. Furthermore, by analysing the explicit form of the $F_{g,n}$ in the dual heterotic theory on $K3\times T^2$, we obtained perfect agreement with the weak coupling limit of (\ref{EqLocalLim}). An interesting open question concerns the study of explicit examples of Calabi-Yau geometries and the analysis of the geometric implications of the consistency conditions derived in this work.

As was also noted in \cite{Huang:2010kf, Huang:2011qx}, the differential equations are not sufficient to define the partition function of the free energy of the topological string since it must be supplemented by suitable boundary conditions. One such condition is the point particle limit in which the topological free energy, when expanded around a point of enhanced gauge symmetry, should reproduce the partition function for $\mathcal{N}=2$ supersymmetric gauge theories in a general $\Omega$-background. In the case of the string couplings $F_{g,n}$, this limit was analysed perturbatively and non-perturbatively in \cite{Antoniadis:2013bja,Antoniadis:2013mna} for $A_\mu^\star$ being identified with the vector superpartner of the heterotic $\bar{T}$-modulus of $T^2$, and indeed the full gauge theory partition function was reproduced. In this work we have extended this analysis and found that all couplings $F_{g,n}$ with $\phi_\star\in \frac{O(2,10)}{O(2)\times O(10)}$ reproduce perturbatively 
Nekrasov's partition function, when expanded around an appropriate point of enhanced gauge symmetry in the string moduli space.

In summary, the findings of this paper further corroborate our proposal  that the string scattering amplitudes $F_{g,n}$ can provide a worldsheet description of the refined topological string. Indeed, we have elucidated the conditions under which such an identification is possible. We have also shown that our proposal is compatible with other approaches towards the refined topological string. In particular, starting only from physical quantities (\emph{i.e.} string scattering amplitudes), we have proposed a way of finding a generalised holomorphic anomaly equation, which \emph{e.g.} in \cite{Huang:2010kf, Huang:2011qx} was postulated as the definition of the refined topological string.

%%%%%%%%%%%%%%%%%
\section*{Acknowledgements}
We thank D.~Orlando, S.~Reffert, S.~Shatashvili, E.~Sokatchev and T.~Taylor for very interesting discussions. S.H. would especially like to thank A.~Iqbal, H.~Jockers, A.~Klemm and Soo-Jong Rey for very inspiring and useful conversations and discussions. I.F. would like to thank the ICTP Trieste and A.Z.A. wishes to thank the CERN Theory Division for their warm hospitality during several stages of this work. S.H. would like to thank the ICTP Trieste, the Asia Pacific Center for Theoretical Physics and Seoul National University for warm hospitality and for creating a stimulating research environment while part of this work was done. The work of S.H. is partly supported by the BQR Accueil EC 2015.

%%%%%%%%%%%%%%%%%%%%%%%%%%%%%%%%%%%%%%%%%%%%%%%%%%%%%%%%%%%%

\appendix
\section{World-Sheet Superconformal Field Theory}\label{App:ConformalAlgebra}

\subsection{\texorpdfstring{The $\mathcal{N}=2$ Superconformal Algebra}{The N=2 Superconformal Algebra}}
The two-dimensional $\mathcal{N}=2$ superconformal algebra of central charge $c$ is spanned by the energy momentum tensor $T$, two supercurrents $G^{\pm}$ and a $U(1)$ Kac-Moody current $J$. The conformal dimensions and the charges of all operators under $J$ are summarised in the following table.
\begin{center}
\begin{tabular}{ccc}\hline
\textbf{operator} & \textbf{conf. weight} & \textbf{$U(1)$}\\ \hline
&&\\[-12pt]
$T$  & $2$ & $0$\\
$G^\pm$ & $3/2$ & $\pm1$\\
$J$ & 1 & 0\\\hline
\end{tabular}
\end{center}
The algebra is realised through the OPE relations among the different operators, which are
{\allowdisplaybreaks
\begin{align}
&T(z)T(w)=\frac{c}{2(z-w)^4}+\frac{2T(w)}{(z-w)^2}+\frac{\partial_wT(w)}{z-w}\,,&&T(z)G^{\pm}(w)=\frac{3G^\pm(w)}{2(z-w)^2}+\frac{\partial_wG^\pm(w)}{z-w}\,,\nonumber\\
&T(z)J(w)=\frac{J(w)}{(z-w)^2}+\frac{\partial_wJ(w)}{z-w}\,,&&J(z)G^\pm(w)=\pm\frac{G^\pm(w)}{z-w}\,,\nonumber\\
&J(z)J(w)=\frac{c}{3(z-w)^2}\,,&&G^+(z)G^+(w)=G^-(z)G^-(w)=0\,,\nonumber
\end{align}}
\begin{align}
G^+(z)G^-(w)=\frac{2c}{3(z-w)^3}+\frac{2J(w)}{(z-w)^2}+\frac{2T(w)+\partial_wJ(w)}{z-w}\,.\label{N2algebra}
\end{align}
Here, we have suppressed all regular terms, which are not important for the computations performed in the main part of this paper.

%%%%%%%%%%%%%%%%%%%%%%%%%%%%%%%%%%%%%%
\subsection{Topological Twist}
In this work, we study correlators in a topologically twisted version of the worldsheet theory discussed above. There are two independent ways to redefine the energy-momentum tensor, which are known as the A- and the B-twist:
\begin{align}
&\text{A-twist}:&&T\to T-\frac{1}{2}\partial J\,,&&\tilde{T}\to \tilde{T}+\frac{1}{2}\bar{\partial}\tilde{J}\,,\\
&\text{B-twist}:&&T\to T-\frac{1}{2}\partial J\,,&&\tilde{T}\to \tilde{T}-\frac{1}{2}\bar{\partial}\tilde{J}\,.
\end{align}
These twists have the effect of shifting the dimensions of all operators by (half of) their charge as shown in the table below.
\begin{center}
\begin{tabular}{c|c|c}\hline
\textbf{operator} & \textbf{A-twisted dimension} & \textbf{B-twisted dimension}\\\hline
&&\\[-12pt]
$T$ & $(2,0)$ & $(2,0)$\\
&&\\[-12pt]
$\tilde{T}$ & $(0,2)$ & $(0,2)$\\\hline 
&&\\[-12pt]
$G^+$ & $(1,0)$ & $(1,0)$\\
&&\\[-12pt]
$\tilde{G}^+$ & $(0,2)$ & $(0,1)$\\\hline 
&&\\[-12pt]
$G^-$ & $(2,0)$ & $(2,0)$\\
&&\\[-12pt]
$\tilde{G}^-$ & $(0,1)$ & $(0,2)$\\\hline 
&&\\[-12pt]
$J$ & $(1,0)$ & $(1,0)$\\
&&\\[-12pt]
$\tilde{J}$ & $(0,1)$ & $(0,1)$\\[1pt]\hline 
\end{tabular}
\end{center}
${}$\\
With these dimensions, we can identify the operators $(G^+,\tilde{G}^-)$ with the left- and right-moving BRST operators in the A-twisted theory, and $(G^+,\tilde{G}^+)$ with the left- and right-moving BRST operators in the B-twisted theory. Physical states of the A- and B-type topological theory are defined to lie in the cohomology of the corresponding BRST operators. Similarly, the operators $(G^-,\tilde{G}^+)$ in the A-twisted model and $(G^-,\tilde{G}^-)$ in the B-twisted model have the right dimensions to be identified with the anti-ghost operators. Indeed, they have the right dimensions to be sewed with the Beltrami-differentials of a Riemann surface, thus providing an integral measure for the twisted correlators as defined in (\ref{TwistAmplitude}).
%%%%%%%%%%%%%%%%%%%%%%%%%%%%%%%%%%%%%%
\subsection{Chiral Ring}\label{App:DefsChiralRing}
In this section, we briefly review Section 2 of \cite{Bershadsky:1993cx}. Our starting point is the chiral ring of the (twisted) worldsheet theory of the Calabi-Yau compactification involving the chiral primary states which satisfy 
\begin{align}
\phi_I\phi_J={C_{IJ}}^K\phi_K+[Q,\cdot ]\,.
\end{align}
Here, our convention for the indices is the same as in the bulk of the paper: the index $I$ ($\bar{I}$) runs over all the (anti-)chiral primaries of the theory. When discussing specific correlation functions in Section~\ref{Sec:TypeII}, we single out one primary field (denoted by $\star$) and label the remaining elements of the (anti-)chiral ring by $i$ ($\bi$) respectively.  

Given the chiral ring, we can define ground states of the theory by acting on a canonical vacuum state $|0\rangle$. Specifically, we have
\begin{align}
|I\rangle=\phi_I|0\rangle+Q|\cdot\rangle\,.\label{DefStates}
\end{align}
Geometrically, this corresponds to inserting the state $\phi_I$ on a hemisphere and attaching an infinitely long cylinder to the boundary. There are two types of measures on this space of states, which are referred to as the topological metric $\eta$ and the hermitian metric $g$:
\begin{align}
&\eta_{IJ}=\langle J|I\rangle\,,&&g_{I\bar{J}}=\langle \bar{J}|I\rangle\,.
\end{align}
The structure of these states generically changes under local deformations of the form
\begin{align}
\Delta S=t_I\int \oint G^-\oint \tilde{G}^\pm \phi_I+\bar{t}_{\bar{I}}\int \oint G^+\oint \tilde{G}^\mp\bar{\phi}_{\bar{I}}\,,
\end{align}
where we introduced the deformation moduli $(t_I,\bar{t}_{\bar{I}})$. This structure takes the form of a bundle, which is usually refered to as the \emph{vacuum bundle} $\mathcal{L}$ of the theory with a base point defined by $|0 \rangle$ and a choice of a base point $(t_0,\bar{t}_0)$. 
%%%%%%%%%%%%%%%%%%%%%%%%

%%%%%%%%%%%%%%%%%%%%%%%%
\section{Boundary Contributions in Type II}\label{App:Boundary}
In this appendix, we consider explicitly the boundary contributions $\mathcal{C}_{\star}^{\text{bdy}}$ and $\mathcal{C}_{\bi}^{\text{bdy}}$ to the type II equations (\ref{DiffEquNoLimStar}) and (\ref{DiffEquNoLim}). As already mentioned, we do not discuss the collision of two punctures which gives rise to curvature dependent contributions.
\subsection{\texorpdfstring{Contribution $\mathcal{C}^{\text{bdy}}_\star$}{Contribution Cbdy}}
We begin with eq.~(\ref{DiffEquNoLimStar}) and consider the boundary contribution
\begin{align}
\mathcal{C}^{\text{bdy}}_\star=\int_{\mathcal{M}_{g,n}}\left\langle\sum_{r=1}^{3g-3+n}\prod_{k\neq r}|\mu_k\cdot G^-|^2\,(\mu_r\cdot T)(\bar{\mu}_r\cdot\tilde{G}^{\pm})\,\left(\int \ins\right)^n\left(\hins\right)^{n}\,\left(\int \oint \tilde{G}^\mp \ins\right)\right\rangle_{\text{twist}}\,.
\end{align}
Besides the collision of punctures (which we neglect), the boundary components contributing to this expression correspond to either a dividing geodesic or a handle degenerating into an infinitely long and thin tube. Even though $\mathcal{C}^{\text{bdy}}_\star$ only has a left-moving energy momentum tensor sewed with the Beltrami differentials (and not a right moving one as well), at a generic point in the moduli space it only receives contributions when one of the operator insertions is integrated over the tube.\footnote{The reason is, that at a generic point in the moduli space, only states with charges $(\pm1,\mp1)$ are massless. Indeed, in order to yield a non-trivial contribution with no additional insertion on the tube, we would require the existence of massless primary states with charges $(+1,\mp 2)$ in the worldsheet theory, which are generically not present.} In order to balance all background charges, the only choice for this operator is $\oint \tilde{G}^{\mp}\bar{\phi}_\star$. 

Furthermore, we can separate $\mathcal{C}^{\text{bdy}}_{\star}$ into the contribution of pinching a handle or a dividing geodesic:
\begin{align}
\mathcal{C}^{\text{bdy}}_\star=\mathcal{B}_{\star}^{\text{geo}}+\mathcal{B}_{\star}^{\text{handle}}\,.
\end{align}
Here $\mathcal{B}_{\star}^{\text{geo}}$ comes from the degeneration of the Riemann surface into two surfaces of lower genera connected by an infinitely long and thin tube:
\begin{align}
&\mathcal{B}^{\text{geo}}_\star=\frac{1}{2}{C_{\inde}}^{JK}{\sum_{g',n'}}^{'}  \int_{\mathcal{M}_{g',n'}}\left\langle\prod_{\ell=1}^{3g'-3+n'}|\mu_\ell\cdot G^-|^2\,\left(\int \ins \right)^{n'} \,\left(\hins \right)^{n'}\,\oint G^-\tilde{G}^\pm\phi_{J} \right\rangle_{\text{twist}}\nonumber\\
&\hspace{2cm}\times \int_{\mathcal{M}_{g-g',n-n'}}\left\langle\prod_{\ell=1}^{3(g-g')-3+n-n'}|\mu_\ell\cdot G^-|^2\,\left(\int \ins \right)^{n-n'}\left(\hins \right)^{n-n'}\,\oint G^-\tilde{G}^\pm\phi_{K} \right\rangle_{\text{twist}}\,,\label{App:Bgeo}
\end{align}
where we used the same notation as in (\ref{NotationThree}), \emph{i.e.} $\sum_{g',n'}^{'} $ excludes summation over the terms $(0,0)$, $(0,1)$, $(g,n-1)$ and $(g,n)$. In particular the exclusion of the terms $(0,1)$ and $(g,n-1)$ is a consequence of the fact that there are no tree-level contact terms between two $\bar{F}_\star$-vector fields. This was explained in \cite{Antoniadis:2010iq} to be a necessary condition to formulate the $F_{g,n}$ as twisted world-sheet correlators like in eq.~(\ref{TwistAmplitude}). The insertions of the form $\oint G^-\tilde{G}^-\phi$ in (\ref{App:Bgeo}) can be interpreted as K\"ahler covariant derivatives:
\begin{align}
\mathcal{B}_{\star}^{\text{geo}}=&\frac{1}{2} {C_{\inde}}^{JK}  {\sum_{g',n'}}^{'}\mc{D}_J F_{g',n'} \mc{D}_K F_{g-g',n-n'}\,.
\end{align}
On the other hand, $\mathcal{B}_{\star}^{\text{handle}}$ captures the contribution of one of the handles degenerating into an infinitely long and thin tube:
\begin{align}
&\mathcal{B}_{\star}^{\text{handle}}=\frac{1}{2}{C_{\inde}}^{JK} \int_{\mathcal{M}_{g-1,n}}\left\langle\prod_{\ell=1}^{3(g-2)+n}|\mu_\ell\cdot G^-|^2\,\left(\int \ins \right)^{n} \,\left(\hins \right)^n\,\oint G^-\tilde{G}^\pm\phi_{J} \oint G^-\tilde{G}^\pm\phi_{K} \right\rangle_{\text{twist}}\nonumber\,.
\end{align}
As before, the insertions $\oint G^-\tilde{G}^{\pm}\phi$ can be interpreted as K\"ahler covariant derivatives:
\begin{align}
\mathcal{B}_{\star}^{\text{handle}}=\frac{1}{2}{C_{\inde}}^{JK} \mc{D}_J \mc{D}_K F_{g-1,n}\,.
\end{align} 

%%%%%%%%%%%%%%%%%%%%%
\subsection{\texorpdfstring{Contribution $\mathcal{C}^{\text{bdy}}_{\bi}$}{Contribution Cbdy}}
We now consider the contribution $\mathcal{C}^{\text{bdy}}_{\bi}$ defined in (\ref{BoundaryDetail}), which can be written as
\begin{align}
\mathcal{C}_{\bi}^{\text{bdy}}=\mathcal{B}_{\bi }^{\text{geo}}+\mathcal{B}_{\bi }^{\text{handle}}
\end{align}
by separating out the boundary components corresponding to the the pinching of a dividing geodesic and the degeneration of a handle. In the following, we compute these contributions explicitly.
\subsubsection{Dividing Geodesic}
$\mathcal{B}_{\bi }^{\text{geo}}$ comes from the degeneration of the Riemann surface into two surfaces of lower genus connected by an infinitely long and thin tube. In order to work it out, we have to distinguish between the contribution stemming from the first line ($\mathcal{B}_{\bi }^{\text{geo},1}$) in (\ref{BoundaryDetail}) and the last two lines ($\mathcal{B}_{\bi }^{\text{geo},2}$). Starting with the former and following the discussion of \cite{Bershadsky:1993cx}, since the insertion of the energy-momentum tensor in (\ref{StressTensorInsertion}) is with respect to the left- and right movers, the expression is in fact a double derivative in the moduli parametrising the surface in the vicinity of the degeneration limit. Therefore, the only non-vanishing contribution arises when one of the operator insertions is integrated over the long-thin tube connecting the two surfaces. We assume that we are at a point in the string moduli space where only anti-chiral fields of charge $(-1,\mp1)$ and dimension $(1,1)
$ become massless and can therefore propagate on the tube. Their contribution can be written in the form
\begin{align}
&\mathcal{B}_{\bi }^{\text{geo},1}=\frac{1}{2}{C_{\bi}}^{JK}{\sum_{g',n'}}^{'}\int_{\mathcal{M}_{g',n'}}\left\langle\prod_{\ell=1}^{3g'-3+n'}|\mu_\ell\cdot G^-|^2\,\left(\int \ins \right)^{n'}\left(\hins \right)^{n'}\,\oint G^-\tilde{G}^\pm\phi_{J}\right\rangle_{\text{twist}}\nonumber\\
&\hspace{2cm}\times \int_{\mathcal{M}_{g-g',n-n'}}\left\langle\prod_{\ell=1}^{3(g-g')-3+n-n'}|\mu_\ell\cdot G^-|^2\,\left(\int \ins \right)^{n-n'}\left(\hins \right)^{n-n'}\,\oint G^-\tilde{G}^\pm\phi_{K} \right\rangle_{\text{twist}}\nonumber\\
&+\frac{n}{2}{C_{\inde}}^{JK}{\sum_{g',n'}}^{'} \int_{\mathcal{M}_{g',n'}}\left\langle\prod_{\ell=1}^{3g'-3+n'}|\mu_\ell\cdot G^-|^2\,\left(\int \ins \right)^{n'-1} \int \bar{\phi}_{\bi }\,\left(\hins \right)^{n'}\,\oint G^-\tilde{G}^\pm\phi_{J} \right\rangle_{\text{twist}}\nonumber\\
&\hspace{2cm}\times \int_{\mathcal{M}_{g-g',n-n'}}\left\langle\prod_{\ell=1}^{3(g-g')-3+n-n'}|\mu_\ell\cdot G^-|^2\,\left(\int \ins \right)^{n-n'}\left(\hins \right)^{n-n'}\,\oint G^-\tilde{G}^\pm\phi_{K} \right\rangle_{\text{twist}}\,,\nonumber
\end{align} 
where we used the notation (\ref{NotationThree}). The insertions of the form $\oint G^-\tilde{G}^-\phi$ can be interpreted as K\"ahler covariant derivatives:
\begin{align}
\mathcal{B}_{\bi }^{\text{geo},1}=&\frac{1}{2}{C_{\bi}}^{JK} {\sum_{g',n'}}^{'}\mc{D}_J F_{g',n'} \mc{D}_KF_{g-g',n-n'}+\frac{n}{2} {C_{\inde}}^{JK}  {\sum_{g',n'}}^{'}\mc{D}_J F_{g',n',\bi } \mc{D}_K F_{g-g',n-n'}\,,
\end{align}
with $F_{g,n,\bi}$ being introduced in (\ref{TwistAmplitudeFi}). 

The contribution of the last two lines of (\ref{BoundaryDetail}), \emph{i.e.} $\mathcal{B}_{\bi }^{\text{geo},2}$, is similar, except for the fact that, due to charge conservation, only $\oint \tilde{G}^{\mp}\bar{\phi}_\star$ and $\oint G^+\bar{\phi}_\star$ can propagate. Therefore, their contribution is
\begin{align}
\mathcal{B}_{\bi }^{\text{geo},2}=&-n\, {C_{\inde}}^{JK}  {\sum_{g',n'}}^{'}\mc{D}_J F_{g',n',\bi } \mc{D}_K F_{g-g',n-n'}\,.
\end{align}
%%%%%%%%%%%%%%%%%%%%%%%%%%%%%%%%%%%%%%%%%

\subsubsection{Handle Degeneration}
$\mathcal{B}_{\bi }^{\text{handle}}$ captures the contribution of one of the handles degenerating into an infinitely long and thin tube. Starting again with the contribution to the first term ($\mathcal{B}_{\bi }^{\text{handle},1}$) in (\ref{BoundaryDetail}) and following the same reasoning as in the previous subsection, one of the integrated insertions must be on this tube. Furthermore, assuming that $g>1$, the only remaining states that can propagate along the handle are anti-chiral primary states of charge $(-1,-1)$ and dimension $(1,1)$. Therefore, we obtain the following two contributions:
\begin{align}
&\mathcal{B}_{\bi }^{\text{handle}}=\frac{1}{2}{C_{\bi}}^{JK} \int_{\mathcal{M}_{g-1,n}}\left\langle\prod_{\ell=1}^{3(g-2)+n}|\mu_\ell\cdot G^-|^2\,\left(\int \ins \right)^{n}\left(\hins \right)^n\,\oint G^-\tilde{G}^\pm\phi_{J} \oint G^-\tilde{G}^\pm\phi_{K} \right\rangle_{\text{twist}}\nonumber\\
&+\frac{n}{2}{C_{\inde}}^{JK} \int_{\mathcal{M}_{g-1,n}}\left\langle\prod_{\ell=1}^{3(g-2)+n}|\mu_\ell\cdot G^-|^2\,\left(\int \ins \right)^{n-1} \int \bar{\phi}_{\bi }\,\left(\hins \right)^n\,\oint G^-\tilde{G}^\pm\phi_{J} \oint G^-\tilde{G}^\pm\phi_{K} \right\rangle_{\text{twist}}\nonumber
\end{align}
The insertions of the form $\oint G^-\tilde{G}^-\phi$ can again be interpreted as K\"ahler covariant derivatives:
\begin{align}
\mathcal{B}_{\bi }^{\text{handle},1}=\frac{1}{2}{C_{\bi}}^{JK}\mc{D}_J \mc{D}_K F_{g-1,n}+\frac{n}{2}{C_{\inde}}^{JK} \mc{D}_J \mc{D}_K F_{g-1,n,\bi }\,.
\end{align} 
The contribution of the last two lines of (\ref{BoundaryDetail}) is similar, except for the fact that only $\oint \tilde{G}^{\mp}\bar{\phi}_\star$ and $\oint G^+\bar{\phi}_\star$ can propagate. Their contribution therefore gives
\begin{align}
\mathcal{B}_{\bi }^{\text{handle},2}=-n\,{C_{\inde}}^{JK} \mc{D}_J \mc{D}_K F_{g-1,n,\bi }\,.
\end{align}
%%%%%%%%%%%%%%%%%%%%%%%%%%%%%%%%%%%%%%%%%

\section{Lattice Momenta}\label{App:Lattice}
In this appendix, we discuss our conventions for the self-dual lattices which are at the heart of heterotic torus compactifications. The basic moduli in the case of $T^2$ are the two-dimensional metric $g_{AB}$, the $B$-field $B_{AB}$ and Wilson-line moduli $W_A^a$. The indices $A,B=1,2$ denote the directions on the torus, while $a=1,\ldots,8$. An explicit parametrisation is given by
\begin{align}
&g_{AB}=\frac{T_2-\frac{W_2^\mu W_2^\mu}{2U_2}}{U_2}\left(\begin{array}{cc} 1 & U_1 \\ U_1 & U_1^2+U_2^2 \end{array}\right)\,&&\text{and} && B_{AB}=\left(T_1-\frac{W_1^\mu W_2^\mu}{2U_2}\right)\left(\begin{array}{cc} 0 & 1 \\ -1 & 0 \end{array}\right)\,,\label{MetricB}
\end{align} 
where we have used the physical moduli
\begin{align}
&T=T_1+iT_2\,,&&U=U_1+iU_2\,,&&W^a=V^a_2-UV^a_1\,.
\end{align} 
Using these objects, we can define the lattice momenta of the $\Gamma^{2,10}$ self-dual lattice as
\begin{align}
P_L^A&=m^A+V^A_a b^a+\tfrac{1}{2}V^A_a V^B_a n_B+B^{AB}n_B+g^{AB}n_B\,,\label{LeftMom}\\
\vec{P}_R&=\left(\begin{array}{c}P^a_R \\ P_R^A\end{array}\right)
=\left(\begin{array}{c}b^a+V^a_An^A \\ m^A+V^A_a b^a+\tfrac{1}{2}V^A_a V^B_a n_B+B^{AB}n_B-g^{AB}n_B\end{array}\right)\,,\label{RightMom}
\end{align}
where $n^a\,,m^a\,,b^a$ are integer numbers. These momenta satisfy the relation
\begin{align}
\tfrac{1}{2}\left(P_L^Ag_{AB}P_L^B-P_R^Ag_{AB}P_R^B-P_R^a P_R^a\right)=2(m_1n_1+m_2n_2)-b^a b^a\,.\label{Discriminant}
\end{align}
For most of the computations carried out in Section~\ref{Sect:HetRealisation}, it is useful to work in a complex basis, \emph{i.e.} instead of $(P_L^A;P_R^A,P_R^a)$ we introduce $(P_L,\bar{P}_L;P_R,\bar{P}_R,P_R^a)$. In order to save writing, we also introduce the shorthand notation
\begin{align}
\xi=\sqrt{(T-\bar T)(U-\bar U)-\frac{1}{2}(W-\bar W)^2}\,,\label{XiDef}
\end{align}
as well as 
\begin{align}
&K_{g,n} \equiv \tau_2^{2g+2n-3}\,\left(\frac{P_L}{\xi}\right)^{2g-2}\left(\frac{P_R}{\xi}\right)^{2n}\,\hat\Gamma^{(2,10)}\,,&&\text{with}&&\hat\Gamma^{2,10}=q^{|P_L|^2}\bar q^{|P_R|^2+\frac{1}{2}p^2}\,.
\end{align}
After some algebra, one can show the following identities:
\begin{align}
 \partial_{\bar T}\hat\Gamma_{2,10}&=-\frac{4\pi\tau_2(U-\bar U)}{\xi^2}\,\bar{P}_L\,P_R\,\hat\Gamma^{2,10}\,,\nonumber\\
 \partial_{\bar U}\hat\Gamma_{2,10}&=-\frac{4\pi\tau_2}{\xi^2(U-\bar U)}\left[\frac{1}{2}(W^a-\bar{W}^a)^2\,\bar P_L\, P_R+\xi^2\,\bar P_L\,\bar P_R+\xi(W^a-\bar{W}^a)\,P^a_R\,\bar P_L\right]\hat\Gamma^{2,10}\,,\nonumber\\
(\partial_{\bar W})^a\hat\Gamma_{2,10}&=\frac{4\pi\tau_2}{\xi^2}\left[(W^a-\bar{W}^a)\,\bar P_L\,P_R+\xi\, P_R^a\,\bar P_L\right]\hat\Gamma^{2,10}\,.\label{WbarLattice}
\end{align}
These allow us to prove that the action of anti-holomorphic derivatives on $K_{g,n}$ is related to that of holomorphic derivatives on $K_{g-1,n}$ up to terms suppressed in the large $T_2$ limit\footnote{In fact, eq.~(\ref{RelKgn}) is exact except for $\bi=\bar{U}$.}:
\begin{align}
	\mathcal{D}_{\bi}K_{g,n}\big|_{T_2\to\infty}= -\frac{1}{2\pi i}\,e^{2K}\,C_{\bi\bj\bar S}\,G^{\bj j}\,\partial_\tau\left(\tau_2^2\,\mathcal{D}_j K_{g-1,n}\right)|_{T_2\to\infty}\,.\label{RelKgn}
\end{align}
where $\mathcal{D}_{\bi}$ is a suitable K\"ahler covariant derivative taking into account the weight of $K_{g,n}$.

%%%%%%%%%%%%%%%%%%%%%%%%%%%

\bibliographystyle{unsrt}

\vfill\eject

\end{document}